\documentclass[acmsmall, nonacm]{acmart}

\AtBeginDocument{%
  \providecommand\BibTeX{{%
    \normalfont B\kern-0.5em{\scshape i\kern-0.25em b}\kern-0.8em\TeX}}}

\setcopyright{acmcopyright}
\copyrightyear{0000}
\acmYear{0000}
\acmDOI{00.0000/0000000.0000000}

\acmJournal{JACM}
\acmVolume{00}
\acmNumber{0}
\acmArticle{000}
\acmMonth{12}

\usepackage{amsthm}
\usepackage{dsfont} 
\usepackage{paralist} 
\usepackage{url} 
\usepackage{pdflscape} 
\usepackage{ragged2e} 
\usepackage{mathtools}

\newtheorem{theorem}{Theorem}
\newtheorem{lemma}[theorem]{Lemma}
\newtheorem{corollary}[theorem]{Corollary}
\newtheorem{definition}{Definition}
\newtheorem{remark}{Remark} 

\usepackage[ruled,vlined]{algorithm2e}

\usepackage{enumerate} 

\usepackage{graphicx}
\usepackage[outdir=./]{epstopdf}
\usepackage{subcaption}

\usepackage{tabulary}
\usepackage{booktabs}

\newcommand{\calBlue}{\mathcal{B}} 
\newcommand{\calRed}{\mathcal{R}} 
\newcommand{\graph}{G} 
\newcommand{\nodeSet}{V} 
\newcommand{\edgeSet}{E} 

\newcommand{\redBirthProb}{r} 

\newcommand{\probEventOne}{p} 
\newcommand{\probEventTwo}{q} 

\newcommand{\deltaboth}{\delta} 
\newcommand{\deltaIn}{\deltaboth_{i}} 
\newcommand{\deltaOut}{\deltaboth_{o}} 

\newcommand{\homophilyRedAnyEvent}{\rho_{\calRed}} 
\newcommand{\homophilyBlueAnyEvent}{\rho_{\calBlue}}  

\newcommand{\homophilyRedEventOne}{\homophilyRedAnyEvent^{(1)}} 
\newcommand{\homophilyBlueEventOne}{\homophilyBlueAnyEvent^{(1)}}  

\newcommand{\homophilyRedEventTwo}{\homophilyRedAnyEvent^{(2)}} 
\newcommand{\homophilyBlueEventTwo}{\homophilyBlueAnyEvent^{(2)}}  

\newcommand{\homophilyRedEventThree}{\homophilyRedAnyEvent^{(3)}} 
\newcommand{\homophilyBlueEventThree}{\homophilyBlueAnyEvent^{(3)}}  

\newcommand{\matrixAnyEvent}{E} 
\newcommand{\matrixEventOne}{\matrixAnyEvent_{1}} 
\newcommand{\matrixEventTwo}{\matrixAnyEvent_{2}} 
\newcommand{\matrixEventThree}{\matrixAnyEvent_{3}} 

\newcommand{\inDegree}{d_{i}} 
\newcommand{\outDegree}{d_{o}} 
\newcommand{\degree}{d} 

\newcommand{\timeValue}{t} 

\newcommand{\thetaIn}{\theta^{(i)}} 
\newcommand{\thetaOut}{\theta^{(o)}} 
\newcommand{\thetaVec}{\theta} 

\newcommand{\nonLinFunction}{F}  

\newcommand{\numNodesIndegreeTime}{m_{k}^{(i)}} 
\newcommand{\numNodesOutdegreeTime}{m_{k}^{(o)}} 
\newcommand{\numNodesIndegreeminusoneTime}{m_{k-1}^{(i)}} 
\newcommand{\numNodesIndegreeminusoneLimit}{M_{k-1}^{(i)}} 

\newcommand{\numNodesIndegreeLimit}{M_{k}^{(i)}} 
\newcommand{\numNodesOutdegreeLimit}{M_{k}^{(o)}} 

\newcommand{\exponentInConstant}{\alpha_{i}} 
\newcommand{\exponentOutConstant}{\alpha_{o}} 
\newcommand{\groupSize}{n} 

\newcommand{\poneBB}{{p}^{(1)}_{\calBlue_{\mathrm{new}}\rightarrow\calBlue_{\mathrm{old}}}} 
\newcommand{\poneBR}{{p}^{(1)}_{\calBlue_{\mathrm{new}}\rightarrow\calRed_{\mathrm{old}}}} 
\newcommand{\poneRR}{{p}^{(1)}_{\calRed_{\mathrm{new}}\rightarrow\calRed_{\mathrm{old}}}} 
\newcommand{\poneRB}{{p}^{(1)}_{\calRed_{\mathrm{new}}\rightarrow\calBlue_{\mathrm{old}}}} 

\newcommand{\ptwoBB}{{p}^{(2)}_{\calBlue_{\mathrm{new}}\leftarrow\calBlue_{\mathrm{old}}}} 
\newcommand{\ptwoBR}{{p}^{(2)}_{\calBlue_{\mathrm{new}}\leftarrow\calRed_{\mathrm{old}}}} 
\newcommand{\ptwoRR}{{p}^{(2)}_{\calRed_{\mathrm{new}}\leftarrow\calRed_{\mathrm{old}}}} 
\newcommand{\ptwoRB}{{p}^{(2)}_{\calRed_{\mathrm{new}}\leftarrow\calBlue_{\mathrm{old}}}} 

\newcommand{\pthreeBB}{{p}^{(3)}_{\calBlue_{\mathrm{old}}\rightarrow \calBlue_{\mathrm{old}}}} 
\newcommand{\pthreeBR}{{p}^{(3)}_{\calBlue_{\mathrm{old}}\rightarrow \calRed_{\mathrm{old}}}} 
\newcommand{\pthreeRR}{{p}^{(3)}_{\calRed_{\mathrm{old}}\rightarrow \calRed_{\mathrm{old}}}} 
\newcommand{\pthreeRB}{{p}^{(3)}_{\calRed_{\mathrm{old}}\rightarrow\calBlue_{\mathrm{old}}}} 

\newcommand{\barponeBR}{\bar{p}^{(1)}_{\calBlue_{\mathrm{new}}\rightarrow\calRed_{\mathrm{old}}}} 
\newcommand{\barponeRR}{\bar{p}^{(1)}_{\calRed_{\mathrm{new}}\rightarrow\calRed_{\mathrm{old}}}}

\newcommand{\barptwoBR}{\bar{p}^{(2)}_{\calBlue_{\mathrm{new}}\leftarrow\calRed_{\mathrm{old}}}} 
\newcommand{\barptwoRR}{\bar{p}^{(2)}_{\calRed_{\mathrm{new}}\leftarrow\calRed_{\mathrm{old}}}}

\newcommand{\barpthreeBR}{\bar{p}^{(3)}_{\calBlue_{\mathrm{old}}\rightarrow \calRed_{\mathrm{old}}}} 
\newcommand{\barpthreeRR}{\bar{p}^{(3)}_{\calRed_{\mathrm{old}}\rightarrow \calRed_{\mathrm{old}}}} 
\newcommand{\barpthreeRB}{\bar{p}^{(3)}_{\calRed_{\mathrm{old}}\rightarrow\calBlue_{\mathrm{old}}}} 

\newcommand{\pOneBRk}{{p}^{(1)}_{\calBlue_{\mathrm{new}}\rightarrow\left(\calRed_{\mathrm{old}}, k\right)}} 
\newcommand{\pOneRRk}{{p}^{(1)}_{\calRed_{\mathrm{new}}\rightarrow\left(\calRed_{\mathrm{old}}, k\right)}} 
\newcommand{\pThreeBRk}{{p}^{(3)}_{\calBlue_{\mathrm{new}}\rightarrow\left(\calRed_{\mathrm{old}}, k\right)}} 
\newcommand{\pThreeRRk}{{p}^{(3)}_{\calRed_{\mathrm{new}}\rightarrow\left(\calRed_{\mathrm{old}}, k\right)}} 

\newcommand{\A}{A^{(i)}} 

\fancypagestyle{mylandscape}{
	\fancyhf{} 
	\fancyfoot{
		\makebox[\textwidth][r]{
			\rlap{\hspace{.75cm}
				\smash{
					\raisebox{4.87in}{
						\rotatebox{90}{\thepage}}}}}}
}

\begin{document}

\title{A Directed, Bi-Populated Preferential Attachment Model with Applications to Analyzing the Glass Ceiling Effect}


\author{Buddhika Nettasinghe}
\email{dwn26@cornell.edu}
\affiliation{%
	\institution{School of Electrical and Computer Engineering, Cornell University}
}
\author{Nazanin Alipourfard}
\email{nazanina@isi.edu}
\affiliation{%
	\institution{Information Sciences Institute, University of Southern California}
}

\author{Vikram Krishnamurthy}
\email{vikramk@cornell.edu}
\affiliation{%
	\institution{School of Electrical and Computer Engineering, Cornell University}
}

\author{Kristina Lerman}
\email{lerman@isi.edu}
\affiliation{%
	\institution{Information Sciences Institute, University of Southern California}
}


\begin{abstract}
Preferential attachment, homophily and, their consequences such as the glass ceiling effect~(\emph{``the unseen, yet unbreakable barrier that keeps minorities and women from rising to the upper rungs of the corporate ladder, regardless of their qualifications or achievements"}) have been well-studied in the context of undirected networks. However, the lack of an intuitive, theoretically tractable model of a directed, bi-populated~(i.e.,~containing two groups) network with variable levels of preferential attachment, homophily and growth dynamics~(e.g.,~the rate at which new nodes join, whether the new nodes mostly follow existing nodes or the existing nodes follow them, etc.) has largely prevented such consequences from being explored in the context of directed networks, where they more naturally occur due to the asymmetry of links~(e.g.~author-citation graphs, Twitter). To this end, we present a rigorous theoretical analysis of the \emph{Directed Mixed Preferential Attachment} model and, use it to analyze the glass ceiling effect in directed networks. More specifically, we derive the closed-form expressions for the power-law exponents of the in- and out- degree distributions of each group~(minority and majority) and, compare them with each other to obtain insights. In particular, our results yield answers to questions such as: \emph{when does the minority group have a heavier out-degree (or in-degree) distribution compared to the majority group? what factors cause the out-degree distribution of a group to be heavier than its own in-degree distribution? what effect does frequent addition of edges between existing nodes have on the in- and out-degree distributions of the majority and minority groups?}. Such insights shed light on the interplay between the structure~(i.e.,~the in- and out- degree distributions of the two groups) and dynamics~(characterized collectively by the homophily, preferential attachment, group sizes and growth dynamics) of various real-world networks. Finally, we utilize the obtained analytical results to characterize the conditions under which the glass ceiling effect emerge in a directed network. Our analytical results are supported by detailed numerical results.
\end{abstract}

\ccsdesc[500]{Theory of computation~Random network models}

\keywords{directed graphs, preferential-attachment, homophily, heterophily, power-law, random graph models, dynamic graphs, minorities, glass ceiling effect}

\maketitle

\section{Introduction}
Preferential attachment~(tendency of popular individuals to become even more popular)~\cite{barabasi1999emergence} and homophily~(preference of individuals to associate with others who have similar attributes)~\cite{mcpherson2001birds} are two of the most well-known sociological concepts related to social network formation~(e.g.~Facebook, Linkedin, dating networks, etc.). It has been previously shown that preferential attachment and homophily lead to the emergence of various sociological phenomena such as the \emph{Glass Ceiling Effect}~(i.e.,~\emph{``the unseen, yet unbreakable barrier that keeps minorities and women from rising to the upper rungs of the corporate ladder, regardless of their qualifications or achievements"}) in undirected networks~\cite{avin2015homophily}.
However, such sociological phenomena have been relatively less studied in the context of directed networks such as author-citation graphs and Twitter, where the relationship between two entities~(that are represented as nodes in the graph) is asymmetric.\footnote{For example, in Twitter, an individual can follow another individual who may not necessarily reciprocate the relationship (i.e.,~information flow may be unidirectional). Similarly, in an author-citation network, one author may cite the work of another who hasn't cited her work. Thus, directed networks are a natural way for representing networks with such asymmetric relationships between the nodes.} A key reason for this gap in literature has been the lack of an intuitive, theoretically tractable generative model that incorporates sociological phenomena such as homophily and preferential attachment into the growth dynamics~(e.g.~rate at which new nodes join, whether the new nodes mostly form out-going links or incoming links with the existing nodes) of directed networks. 

\begin{figure}
	\centering
	\includegraphics[width=\linewidth]{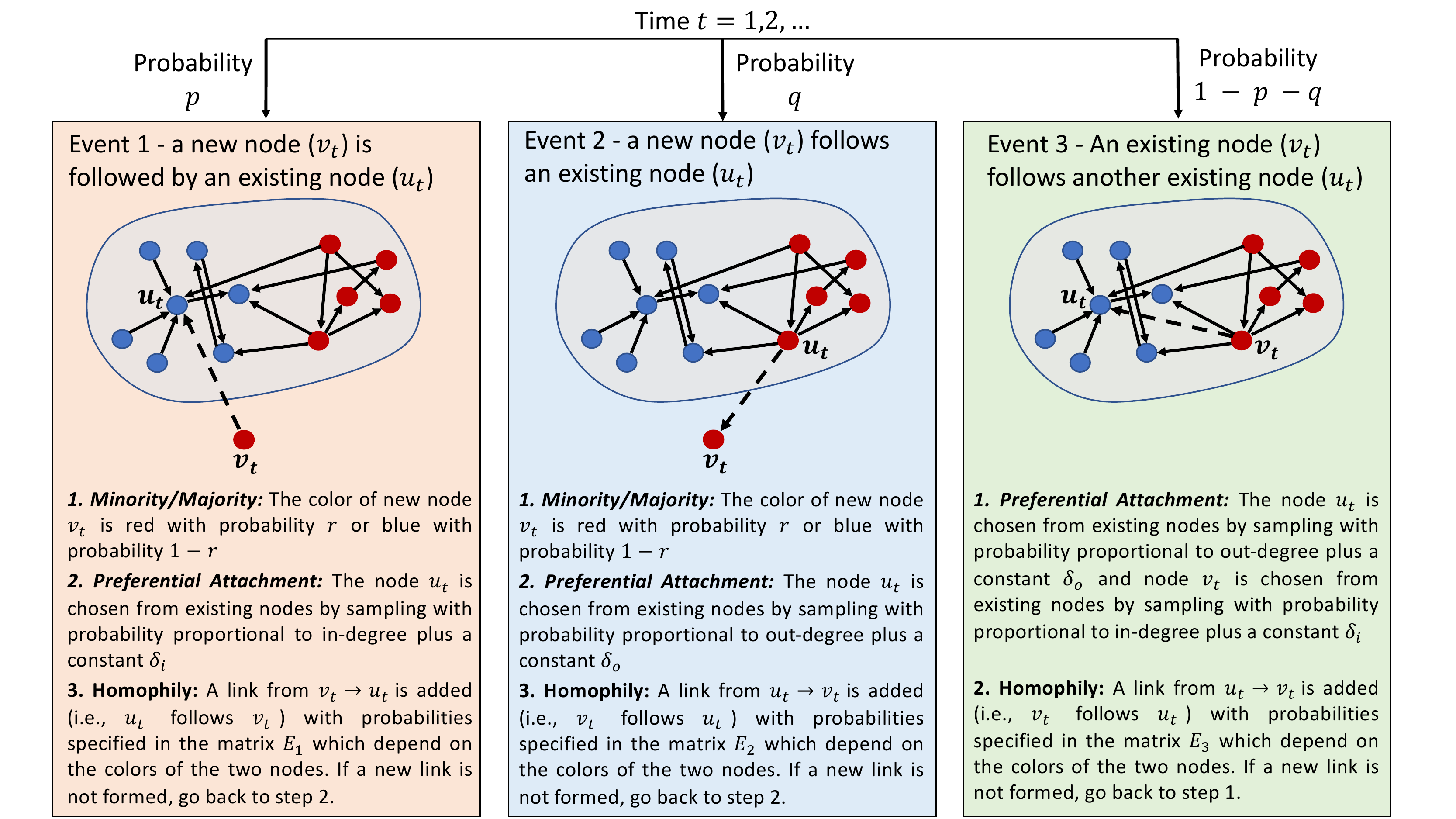}
	\caption{The Directed Mixed Preferential Attachment~(DMPA) model generates a bi-populated directed network via three events that constitute the growth of the directed network~(where, the direction of the links indicate the information flow~e.g.~a link from a node $a$ to to another node~$b$ indicates that information flows from~$a$ to~$b$~i.e.,~$b$ follows $a$). The first two events correspond to the appearance of a new node: a new directed edge is created from a new node to an existing node (Event 1 - probability~$\probEventOne $), a new directed edge is created from an existing node to a new node (Event 2 - probability~$\probEventTwo $). Event 3~(with probability~$1-\probEventOne -\probEventTwo $) corresponds to network densification: an edge is created between two existing nodes. Therefore, varying $\probEventOne, \probEventTwo$ will create different growth dynamics for the graph. The probability $\redBirthProb$ (with which the new node added in the events~1 or~2 belong to the red group) determines whether red group is the minority~($\redBirthProb<0.5$), majority~($\redBirthProb>0.5$) or neither~($\redBirthProb=0.5$). In each event, the existing nodes that are potentially included in the new link are chosen according to directed preferential attachment: existing nodes with larger in-degrees are more likely to follow more nodes and existing nodes with larger out-degrees are more likely to be followed by more nodes. Whether the potential link~(chosen based on preferential attachment) is actually added to the network or not is decided randomly with the probabilities specified in the homophily matrices $\matrixAnyEvent_{1}, \matrixAnyEvent_{2}, \matrixAnyEvent_{3}$. For example, in the event 1 of the example shown in figure, the blue node will follow the red node with probability $1-\homophilyBlueEventOne$. Similarly, in the events 2, 3 of the example shown in figure, the potential links are added with probabilities $\homophilyRedEventTwo$ and $1-\homophilyBlueEventThree$, respectively. In Sec.~\ref{sec:properties} we show that. under the DMPA model, each group~(red, blue) has power-law in- and out- degree distributions asymptotically, present their closed form expressions and, show that they can be used to characterize the conditions that lead to the \emph{glass ceiling effect}.}
	\label{fig:model_diagram}
\end{figure}

Towards this end, in our previous work \cite{nettasinghe2020emergence}, we devised a dynamic model of a directed network that contains two types of nodes~(blue nodes and red nodes) called the \emph{Directed Mixed Preferential Attachment Model} (which is summarized in Sec.~\ref{sec:model} - see also Fig.~\ref{fig:model_diagram}). In this context, the aim of this paper is to provide an in-depth theoretical analysis of the DMPA model, discuss the insight it provides about real-world directed networks and, illustrate how it is useful for studying phenomena such as the glass ceiling effect in directed networks.\footnote{While this paper builds upon our previous work~\cite{nettasinghe2020emergence}, the overlap between the two papers is limited to the DMPA model~(which is summarized in Sec.~\ref{subsec:DMPA_model} of this paper): \cite{nettasinghe2020emergence} focuses on the emergence of\emph{ power-inequality}~(i.e.,~an average based definition of the asymmetry of influence of each group) whereas the current paper derives the closed form expressions for the in- and out- degree distributions of each group under the DMPA model, compares them with each other to obtain insight and, illustrates how they are useful in modeling and analyzing the \emph{glass ceiling effect}~(defined using the tails of the in- and out- degree distributions of each group) in directed networks.}

\vspace{0.2cm}
\noindent
{\bf Main contributions} of this paper are as follows. 
\begin{compactenum}[\hspace{0.5cm}(1.)]
	\item we show that the in- and out- degree distributions of both red and blue groups follow power-laws under the DMPA model and, derive the closed form expressions for the power-law exponents~(Sec.~\ref{sec:convergence}). 
	
	\item we analytically compare the in- and out- degree distributions of the same group and between the two groups under various parameter values (various degrees of the preferential attachment, different group sizes, various homophily levels for each group and different growth dynamics) and discuss the insight that the comparison yields~(Sec.~\ref{subsec:discussion}).
	
	\item we use the DMPA model and its analysis to study the emergence of the glass ceiling effect in directed networks and, provide insight into the conditions that give rise to it~(Sec.~\ref{subsec:GCE}).
\end{compactenum}

The above results shed light on how real-world directed networks~(e.g.~Twitter, author-citation networks, supply chains, etc.) are affected by various sociological phenomena collectively. For example, our results yield answers to key questions such as,
\begin{compactenum}[\hspace{0.5cm}(i.)]
	\item 	\emph{when does the minority group have a heavier out-degree~(or~in-degree) distribution compared to the majority group?}
	
	\item \emph{what factors cause the out-degree distribution of a group to be heavier than its in-degree distribution?}
	
	\item \emph{how does frequently adding new nodes with outgoing~(or incoming) edges affect the in- and out- degree distributions of the majority and minority groups?}
	
	\item \emph{what effect does frequent addition of edges between existing nodes have on the  in- and out- degree distributions of the majority and minority groups?}
	
	\item \emph{how do the group size, degree of preferential attachment, homophily level of each group and growth dynamics affect the glass ceiling effect?}
\end{compactenum}
Via answers to such questions, our results yield insight into the interplay between structure~(i.e.,~the in- and out- degree distributions of the two groups and how they compare with each other) and dynamics~(modeled collectively by homophily, preferential attachment, growth parameters, etc.) of directed networks and thus, help to bridge the knowledge gap between well-studied undirected networks and the relatively less studied directed networks. Therefore, our results might be of use in particular to the practitioners in fields such as network science, computational social science and economics. 

\vspace{0.2cm}
\noindent
{\bf Organization: } Sec.~\ref{sec:model} recaps the \emph{Directed Mixed Preferential Attachment} model and discusses the intuition behind it. Then, Sec.~\ref{sec:convergence} presents the results which show that the in-degree and out-degree distribution of each group~(red, blue) follow a power-law along with their closed-form expressions. Sec.~\ref{sec:properties} compares the inter-group and intra-group degree distributions under various cases and illustrates how the results are useful for analyzing the emergence of the glass ceiling effect in directed networks. 

\vspace{0.2cm}
\noindent
{\bf Related Work: } Preferential-attachment model was first presented in \cite{barabasi1999emergence} for undirected networks and leads to power-law degree distributions. Subsequently, \cite{cooper2003general,bollobas2003directed} extended the idea of preferential-attachment to directed graphs where, the in- and out- degree distributions take power-law form. While the models presented in \cite{cooper2003general,bollobas2003directed} are useful in explaining the power-law degree distributions observed in directed graphs, they are limited to having a homogeneous population~i.e.,~does not contain different types of nodes. Hence, models studied in \cite{cooper2003general,bollobas2003directed} do not allow exploring effects such as homophily that naturally needs multiple types of nodes. In another direction, a preferential attachment model of an undirected graph with two types of nodes has been presented in \cite{avin2015homophily, avin2020mixed}. In particular, \cite{avin2015homophily} has shown that preferential-attachment, homophily and minorities are necessary and sufficient conditions for the emergence of a glass ceiling effect in an undirected graph. However, the model in \cite{avin2015homophily, avin2020mixed} is limited to the setting of an undirected graph and therefore, is not helpful in exploring power-law in- and out- degree distributions and their consequences observed in directed graphs. Further, the model in \cite{avin2015homophily, avin2020mixed} also does not allow adding links between existing nodes or varying the degree of preferential-attachment, which makes it difficult to replicate the dynamics of many real-world networks where edges are frequently formed between existing nodes with varying degrees of preferential attachment.  Towards this end, the Directed Mixed Preferential Attachment~(DMPA) model analyzed in this paper allows us to simultaneously study the collective effects of directed preferential-attachment, homophily as well as the growth dynamics of the network. The analysis of the model is based on tools from stochastic averaging theory~\cite{kushner2003}.

\section{Directed Mixed Preferential Attachment~(DMPA) Model}
\label{sec:model}
This section recaps the Directed Mixed Preferential Attachment~(DMPA) model~\cite{nettasinghe2020emergence} and discusses how it represents various sociological phenomena such as homophily, preferential attachment. In addition, we also discuss how the DMPA model can replicate various growth dynamics observed in real-world social networks.

\subsection{Directed Mixed preferential attachment model $\mathbf{{DMPA}(\redBirthProb, \probEventOne , \probEventTwo, \matrixEventOne , \matrixEventTwo , \matrixEventThree, \deltaIn, \deltaOut)}$} 
\label{subsec:DMPA_model}

The directed mixed preferential attachment model $\mathrm{DMPA}(\redBirthProb, \probEventOne , \probEventTwo, \matrixEventOne , \matrixEventTwo , \matrixEventThree, \deltaIn, \deltaOut)$ generates a bi-populated  random graph containing two types of nodes: red and blue. The model parameters $\matrixAnyEvent _i, i \in \{1,2,3\}$ are  matrices of the form
\begin{equation}
	\matrixAnyEvent_{i} = \begin{bmatrix}
		\homophilyBlueAnyEvent^{(i)} & 1 - \homophilyRedAnyEvent^{(i)} \\
		1 - \homophilyBlueAnyEvent^{(i)} & \homophilyRedAnyEvent^{(i)} 
	\end{bmatrix}
\end{equation} and, $\redBirthProb \in [0,1]$, $\probEventOne, \probEventTwo   \in [0,1]$ such that $\probEventOne + \probEventTwo  \leq 1$ and $\deltaIn, \deltaOut> 0$.

The social network at time $\timeValue$ is denoted by the directed graph $\graph_\timeValue = (\nodeSet_\timeValue, \edgeSet_\timeValue)$ where the set of nodes $\nodeSet_\timeValue$ can be partitioned into two subsets: blue nodes~$\calBlue_\timeValue$ and red nodes~$\calRed_\timeValue$. The initial graph $\graph_0$ is chosen arbitrarily. At every time step $\timeValue = 1, 2, \dots$, one of the following three events take place:
\begin{enumerate}
	\item Event 1 (with probability $\probEventOne$): An existing node following the new node
	\begin{compactenum}[i.]
		\item {\it Minority - majority partition:} A new node $v_\timeValue$  is born and is assigned color red with probability $\redBirthProb$ or blue with probability $1-\redBirthProb$.
		
		\item {\it Preferential attachment in following a new node:} An existing node $u_\timeValue \in V_\timeValue$ is chosen by sampling with probability proportional to $\inDegree(u_\timeValue) + \deltaIn$. 
		
		\item {\it Homophily/heterophily in following a new node:} If both $u_\timeValue, v_\timeValue$ are red (resp. blue), a link $(v_\timeValue, u_\timeValue)$ is formed with probability $\homophilyRedEventOne$ (resp. $\homophilyBlueEventOne$). Otherwise, if $u_k$ is red (resp. blue) and $v_\timeValue$ is blue (resp. red), a link $(v_\timeValue, u_\timeValue)$ is formed with probability $1 - \homophilyRedEventOne$ (resp. $1 - \homophilyBlueEventOne$). 
		
		\item Above steps ii, iii are repeated until an outgoing edge is added to $v_\timeValue$.
	\end{compactenum}

	\item Event 2 (with probability $\probEventTwo$): The new node following an existing node
	\begin{compactenum}[i.]
		\item {\it Minority - majority partition:} A new node $v_\timeValue$  is born and is assigned color red with probability $\redBirthProb$ or blue with probability $1-\redBirthProb$.
		
		\item {\it Preferential attachment in following an existing node:} An existing node $u_\timeValue \in V_\timeValue$ is chosen by sampling with probability proportional to $\outDegree(u_\timeValue) + \deltaOut$ . 
		
		\item {\it Homophily/heterophily in following an existing node:} If both $u_\timeValue, v_\timeValue$ are red (resp. blue), a link $(u_\timeValue, v_\timeValue)$ is formed with probability $\homophilyRedEventTwo$ (resp. $\homophilyBlueEventTwo$). Otherwise, if $u_\timeValue$ is red (resp. blue) and $v_\timeValue$ is blue (resp. red), a link $(u_\timeValue, v_\timeValue)$ is formed with probability $1 - \homophilyBlueEventTwo$ (resp. $1 - \homophilyRedEventTwo$). 
		
		\item Above steps ii, iii are repeated until an incoming edge is added to $v_\timeValue$.
	\end{compactenum}

	\item Event 3 (with probability $1 - \probEventOne  - \probEventTwo$): Edge densification
	\begin{compactenum}[i.]	
		\item {\it Preferential attachment without new nodes:} An existing node $u_\timeValue \in V_\timeValue$ is chosen by sampling with probability proportional to $\outDegree(u_\timeValue) + \deltaOut$ and another node $v_\timeValue \in V_\timeValue$ is chosen by sampling with probability proportional to $\inDegree(u_\timeValue) + \deltaIn$. 
		
		\item {\it Homophily/heterophily without new nodes:} If both $u_\timeValue, v_\timeValue$ are red (resp. blue), a link $(u_\timeValue, v_\timeValue)$ is formed with probability $\homophilyRedEventThree$ (resp. $\homophilyBlueEventThree$). Otherwise, if $u_\timeValue$ is red (resp. blue) and $v_\timeValue$ is blue (resp. red), a link $(u_\timeValue, v_\timeValue)$ is formed with probability $1 - \homophilyBlueEventThree$ (resp. $1 - \homophilyRedEventThree$). 
		
		\item Above steps i, ii are repeated until a new edge is added to the graph.
	\end{compactenum}

\end{enumerate}

\subsection{Intuition behind the DMPA model and the sociological phenomena that it represents} The proposed model $\mathrm{DMPA}(\redBirthProb, \probEventOne , \probEventTwo, \matrixEventOne , \matrixEventTwo , \matrixEventThree, \deltaIn, \deltaOut)$ incorporates many well-known sociological and network scientific phenomena and provides the flexibility to explore the collective effect of those phenomena both analytically as well as using simulations~(as we discuss in the subsequent sections). Let us briefly discuss the intuition behind the DMPA model and how it represents various sociological phenomena. 

In the first event of the DMPA model, the probability $\redBirthProb$ with which a new node is added determines whether red is the minority ($\redBirthProb < 0.5)$ or the majority ($\redBirthProb > 0.5$) asymptotically.  In the step~ii of event~1, a potential follower is chosen for the new node from the existing nodes by sampling proportional to the in-degrees plus a constant $\deltaIn$. The idea is that a node who is already following a lot of people is likely to follow the new node~(i.e.,~preferential attachment)~-~following others according to preferential attachment will give rise to a power-law in-degree distribution as we will prove later. The potential follower then follows the new node based on probabilities that depend on their colors as specified in the matrix $\matrixEventOne$: if both nodes are of same color, the connection probability is specified by the diagonal entry corresponding to their color~(red or blue) whereas if they are different, the connection probability is specified by the off diagonal entry corresponding to the color of the follower node (i.e. the node chosen in step ii). This process is repeated until an edge that points from the new node to an existing node is added. A similar explanation holds for event~2 where the new node follows an existing node~(i.e.,~an edge is added from an existing node to the new node). In the event~3, a random friend (i.e.,~a node sampled proportional to the out-degree plus the constant~$\deltaOut$) and a random follower (i.e.~a node sampled proportional to the in-degree plus the constant~$\deltaIn$) are chosen and an edge (that points from the random friend to the random follower) is added with a probability that depends on their colors as specified by the the matrix~$\matrixEventThree$.

For any one of the three events~$i \in \{1,2,3\}$, if $\homophilyBlueAnyEvent^{(i)} > 0.5$, the blue group is homophilic in the event~$i$~i.e.,~a blue individual in event~$i$ is more likely to follow a blue individual compared to a red individual. On the other hand, if $\homophilyBlueAnyEvent^{(i)} < 0.5$, the blue group is heterophilic in the event~$i$~i.e.,~a blue individual in event~$i$ is more likely to follow a red individual compared to a blue individual. A similar interpretation holds for the red group (with the parameter~$\homophilyBlueAnyEvent^{(i)}, i \in \{1,2,3\}$) as well. 

Thus, the DMPA model allows us to explore the outcomes of many different scenarios by varying the parameters. Some examples are as follows. 
\begin{compactitem}
	\item {\it Group sizes:} We can control the composition of the two groups~(red, blue) in the graph by varying the probability~$\redBirthProb$ with which a new red node is born.
	
	\item {\it Growth dynamics of the graph:} We can vary the  probabilities of the three events that constitute the growth of the network: new node joining and being followed by existing individuals~(event~1), a new node joining and following an existing individual~(event~3) and edge densification~(event~3). By changing these probabilities, we can replicate the growth of various real-world networks. For example, setting $\probEventTwo  > 1-\probEventOne - \probEventTwo > \probEventOne $ corresponds to the case where new nodes joining and following existing individuals~(event~2) is the most likely event to happen at any time with the second likely event being the existing individuals following existing individuals~(event~3 - edge densification) - this setting resembles the author citation network of a new research field where new authors frequently join and cite the existing ones. 
	
	\item {\it Homophily and heterophily level for each event:} We can model how the homophily and heterophily of a node varies depending on the age of two nodes with respect to each other. For example, if $\homophilyBlueEventOne, \homophilyBlueEventTwo, \homophilyBlueEventThree$ are different from each other, it means that an individual of the blue group has different homophily/heterophily level depending on the event. To illustrate this further, let us assume that blue and red correspond to ``elite" vs ``non-elite" academic institutes: Then, an author from an elite university~(author~1) may have different preferences to cite another author~(author~2) depending on whether author~1 is a senior professor and author~2 is a junior graduate student~(event~1) and vice versa~(event~2) or, both author~1 and author~2 are senior professors~(event~3). It is conceivable that for a junior student~(with no prior research record) the affiliation may matter more compared to a senior professor~(who has a lengthy research record). Thus, the ability to vary the three matrices~$\matrixAnyEvent_{i}, i \in \{1,2,3\}$ allows us to explore the outcomes in such situations where the level of homophily/heterophily depends on the seniority of one individual with respect to other that may potentially be connected together via a directed link. 
	
	\item {\it Preferential attachment parameters:} We can vary the importance of the in-degree and out-degree separately using the parameters~$\deltaIn$ and $\deltaOut$ respectively. For example, if $\deltaOut$ is relatively small, then the probability that an existing node gets picked as a candidate to be followed in events~2,~3 depends mostly on the out-degree of that node only. However, if $\deltaOut$ is relatively large, the dependence on the out-degree of that node is reduced. 
\end{compactitem}

\section{Asymptotic power-law form of the degree distributions}
\label{sec:convergence}

In this section, we will analyze the asymptotic properties of the DMPA model using tools from the stochastic averaging theory. The main result of this section~(Theorem~\ref{th:convergence_DMPA}) states that~(under some minor assumptions): \\ 1.~the sum of the in-degrees~(resp.~out-degrees) of red nodes as a fraction of the sum of the in-degrees~(resp.~out-degrees) of all nodes converges almost surely~(with probability~1) to a unique value that does not depend on the initial state of the network \\
2.~the asymptotic in- and out- degree distributions of each group~(red, blue) are  almost surely~(with probability~1) power-laws and, provide closed form expressions for their power-law exponents.\\ 
These results are useful for exploring the collective effect of preferential-attachment, homophily and growth dynamics on the emergence of various interesting properties as we discuss later~(in Sec.~\ref{sec:properties}).

We first need establish the notation that is required to state the main result. Let 
\begin{equation}
	\inDegree(\calRed_\timeValue) = \sum_{v \in{\calRed_\timeValue}}\inDegree(v),\quad  \outDegree(\calRed_\timeValue) = \sum_{v \in{\calRed_\timeValue}}\outDegree(v), \quad \degree_\timeValue = \sum_{v \in V_\timeValue} \inDegree(v).
\end{equation}
and, 
\begin{equation}
	\label{eq:thetaTime}
	\thetaIn_\timeValue = \frac{\inDegree(\calRed_\timeValue)}{\degree_\timeValue}, \quad \thetaOut_\timeValue = \frac{\outDegree(\calRed_\timeValue)}{\degree_\timeValue}.
\end{equation}
Thus, $\inDegree(\calRed_\timeValue), \outDegree(\calRed_\timeValue)$ denote the sum of the in-degrees of all red nodes at time $\timeValue$ and the sum of out-degrees of all red nodes at time $\timeValue$ respectively and, $\degree_\timeValue$ denotes the sum of in-degrees of all nodes in the graph~(which is also equal to the sum of out-degrees of all nodes) at time~$\timeValue$.  Then, $\thetaIn_\timeValue, \thetaOut_\timeValue$ express the sum of the in-degrees~$\inDegree(\calRed_\timeValue)$ and the sum of the out-degrees~$\outDegree(\calRed_\timeValue)$ of red-nodes at time $\timeValue$ as a fraction of sum of the degrees~$\degree_\timeValue$ of all nodes. Further, we use
\begin{equation}
	\label{eq:numNodesDegreeTime}
	\numNodesIndegreeTime(\calRed_\timeValue) = \sum_{v \in{\calRed_\timeValue}} \mathds{1}_{\{\inDegree(v) = k\}}, \quad \numNodesOutdegreeTime(\calRed_\timeValue)= \sum_{v \in{\calRed_\timeValue}} \mathds{1}_{\{\outDegree(v) = k\}}
\end{equation}
to denote the number of red nodes with in-degree $k$ at time $\timeValue$ and the number of red nodes with out-degree $k$ at time $\timeValue$, respectively. Similarly, we use~$\numNodesIndegreeTime(\calBlue_\timeValue), \numNodesOutdegreeTime(\calBlue_\timeValue)$ to denote the analogous quantities for blue nodes.

With the above notation, our main result of this section below establishes the convergence of the properties of the DMPA model. 

\begin{theorem}[Almost sure convergence of the DMPA model and asymptotic degree distributions]
	\label{th:convergence_DMPA}
	Consider the DMPA model $\mathrm{DMPA}(\redBirthProb, \probEventOne , \probEventTwo, \matrixEventOne , \matrixEventTwo , \matrixEventThree, \deltaIn, \deltaOut)$ and assume that $\deltaIn = \deltaOut = \deltaboth$ and,
	\begin{equation}
		\matrixAnyEvent_{1} = \matrixAnyEvent_{2} = \matrixAnyEvent_{3} = \matrixAnyEvent = \begin{bmatrix}
			\homophilyBlueAnyEvent & 1 - \homophilyRedAnyEvent \\
			1 - \homophilyBlueAnyEvent & \homophilyRedAnyEvent \end{bmatrix}.
	\end{equation}
	Then, there exists $\deltaboth^{*} > 0$ such that, for all $\deltaboth > \deltaboth^{*}$, the following statements hold.
	\begin{compactenum}[(i.)]
		\item As time $\timeValue$ tends to infinity, the state of the system $\thetaVec_\timeValue = [\thetaIn_\timeValue, \thetaOut_\timeValue]^T$~(where $\thetaIn_\timeValue, \thetaOut_\timeValue$ are the normalized sum of in-degrees and the normalized sum of out-degrees of red nodes defined in (\ref{eq:thetaTime}) converges to a unique value ${\thetaVec_{*} = [\thetaIn_{*}, \thetaOut_{*}]^T \in [0,1]^2}$ almost surely~(i.e.,~with probability $1$).
	
		\item As time $\timeValue$ tends to infinity, the values $\frac{\numNodesIndegreeTime(\calRed_\timeValue)}{\timeValue}$, $\frac{\numNodesOutdegreeTime(\calRed_\timeValue)}{\timeValue}$ (where $\numNodesIndegreeTime(\calRed_\timeValue), \numNodesOutdegreeTime(\calRed_\timeValue)$ are the sum of in-degrees and sum of out-degrees of red nodes at time $\timeValue$ defined in (\ref{eq:numNodesDegreeTime})) converge almost surely to $\numNodesIndegreeLimit(\calRed), \numNodesOutdegreeLimit(\calRed)$ respectively, where,
		\begin{equation}
			\label{eq:power_law_form}
			\numNodesIndegreeLimit(\calRed) \propto k^{-\left(1+\frac{1}{\exponentInConstant(\calRed)}\right)}, \quad \numNodesOutdegreeLimit(\calRed) \propto k^{-\left(1+ \frac{1}{\exponentOutConstant(\calRed)}\right)}
		\end{equation}
		and,
		\begin{align}
		\begin{split}
		\label{eq:exponentInContant}
		\exponentInConstant(\calRed) =& \frac{\redBirthProb\probEventOne \homophilyRedAnyEvent}{\delta \left(\probEventOne + \probEventTwo\right) - \homophilyBlueAnyEvent \left(\delta \left(1 - \redBirthProb\right) \left(\probEventOne + \probEventTwo\right) - \thetaIn_{*} + 1\right) - \left(1 - \homophilyRedAnyEvent\right) \left(\redBirthProb\delta \left(\probEventOne + \probEventTwo\right) + \thetaIn_{*}\right) + 1} + \\
		&\frac{\probEventOne \left(1 - \redBirthProb\right) \left(1 - \homophilyRedAnyEvent\right)}{\delta \left(\probEventOne + \probEventTwo\right) - \homophilyRedAnyEvent \left(\redBirthProb\delta \left(\probEventOne + \probEventTwo\right) + \thetaIn_{*}\right) - \left(1 - \homophilyBlueAnyEvent\right) \left(\delta \left(1 - \redBirthProb\right) \left(\probEventOne + \probEventTwo\right) - \thetaIn_{*} + 1\right) + 1} + \\
		&\frac{\left(1 -\probEventOne-\probEventTwo \right) \left( \homophilyRedAnyEvent \left(\redBirthProb\delta \left(\probEventOne + \probEventTwo\right) + \thetaOut_{*}\right) + \left(1 - \homophilyRedAnyEvent\right) \left(\delta \left(1 - \redBirthProb\right) \left(\probEventOne + \probEventTwo\right) - \thetaOut_{*} + 1\right) \right) }
		{
			\splitfrac{- \homophilyBlueAnyEvent \left(\redBirthProb\delta \left(\probEventOne + \probEventTwo\right) + \thetaOut_{*}\right) \left(\delta \left(1 - \redBirthProb\right) \left(\probEventOne + \probEventTwo\right) - \thetaIn_{*} + 1\right) }
			{
				\splitfrac{- \homophilyRedAnyEvent \left(\redBirthProb\delta \left(\probEventOne + \probEventTwo\right) + \thetaIn_{*}\right) \left(\delta \left(1 - \redBirthProb\right) \left(\probEventOne + \probEventTwo\right) - \thetaOut_{*} + 1\right)}{ \splitfrac{
 - \left(1 - \homophilyBlueAnyEvent\right) \left(\delta \left(1 - \redBirthProb\right) \left(\probEventOne + \probEventTwo\right) - \thetaIn_{*} + 1\right) \left(\delta \left(1 - \redBirthProb\right) \left(\probEventOne +\probEventTwo\right) - \thetaOut_{*} + 1\right) }{
	- \left(1 - \homophilyRedAnyEvent\right) \left(\redBirthProb\delta \left(\probEventOne + \probEventTwo\right) + \thetaIn_{*}\right) \left(\redBirthProb\delta \left(\probEventOne + \probEventTwo\right) + \thetaOut_{*}\right) + \left(\delta \left(\probEventOne + \probEventTwo\right) + 1\right)^{2}}}
			}
		}
		\end{split}		
		\end{align}
		\begin{align}
		\begin{split}
		\label{eq:exponentOutContant}
		\exponentOutConstant(\calRed) =&	\frac{\redBirthProb\probEventTwo \homophilyRedAnyEvent}{\delta \left(\probEventOne + \probEventTwo\right) - \homophilyRedAnyEvent \left(\delta \left(1 - \redBirthProb\right) \left(\probEventOne + \probEventTwo\right) - \thetaOut_{*} + 1\right) - \left(1 - \homophilyRedAnyEvent\right) \left(\redBirthProb\delta \left(\probEventOne + \probEventTwo\right) + \thetaOut_{*}\right) + 1} + \\
		&\frac{\probEventTwo \left(1 - \redBirthProb\right) \left(1 - \homophilyBlueAnyEvent\right)}{\delta \left(\probEventOne + \probEventTwo\right) - \homophilyBlueAnyEvent \left(\redBirthProb\delta \left(\probEventOne + \probEventTwo\right) + \thetaOut_{*}\right) - \left(1 - \homophilyBlueAnyEvent\right) \left(\delta \left(1 - \redBirthProb\right) \left(\probEventOne + \probEventTwo\right) - \thetaOut_{*} + 1\right) + 1} + \\
		&\frac{\left(1 - \probEventOne - \probEventTwo \right) \left(\homophilyRedAnyEvent \left(\redBirthProb\delta \left(\probEventOne + \probEventTwo\right) + \thetaIn_{*}\right)  + \left(1 - \homophilyBlueAnyEvent\right)  \left(\delta \left(1 - \redBirthProb\right) \left(\probEventOne + \probEventTwo\right) - \thetaIn_{*} + 1\right)\right)}
		{
			\splitfrac{- \homophilyBlueAnyEvent \left(\redBirthProb\delta \left(\probEventOne + \probEventTwo\right) + \thetaOut_{*}\right) \left(\delta \left(1 - \redBirthProb\right) \left(\probEventOne + \probEventTwo\right) - \thetaIn_{*} + 1\right) }{\splitfrac{- \homophilyRedAnyEvent \left(\redBirthProb\delta \left(\probEventOne + \probEventTwo\right) + \thetaIn_{*}\right) \left(\delta \left(1 - \redBirthProb\right) \left(\probEventOne + \probEventTwo\right) - \thetaOut_{*} + 1\right) }{\splitfrac{- \left(1 - \homophilyBlueAnyEvent\right) \left(\delta \left(1 - \redBirthProb\right) \left(\probEventOne + \probEventTwo\right) - \thetaIn_{*} + 1\right) \left(\delta \left(1 - \redBirthProb\right) \left(\probEventOne + \probEventTwo\right) - \thetaOut_{*} + 1\right)}{ - \left(1 - \homophilyRedAnyEvent\right) \left(\redBirthProb\delta \left(\probEventOne + \probEventTwo\right) + \thetaIn_{*}\right) \left(\redBirthProb\delta \left(\probEventOne + \probEventTwo\right) + \thetaOut_{*}\right) + \left(\delta \left(\probEventOne + \probEventTwo\right) + 1\right)^{2}}}
			}
		}		
		\end{split} \,.
		\end{align}
	\end{compactenum}
\end{theorem}

\noindent
{\bf Key idea behind the proof of Theorem~\ref{th:convergence_DMPA}:} The full proof of the Theorem~\ref{th:convergence_DMPA} is given in Appendix~\ref{appendix:proof_convergence_DMPA} - the first part of it has been established in our earlier work in \cite{nettasinghe2020emergence} and, is stated here with additional explanation since second part is dependent on it. Let us look at the high level idea behind the proof in order to understand the usefulness of Theorem~\ref{th:convergence_DMPA}.  The key idea behind the proof of the part~i of Theorem~\ref{th:convergence_DMPA} is that $\thetaVec_\timeValue = [\thetaIn_\timeValue, \thetaOut_\timeValue]^T$ is a Markov process whose evolution can be expressed as a non-linear stochastic dynamical system driven by martingale difference noise~i.e.,~
\begin{equation}
\label{eq:theta_stochastic_approximation}
	\thetaVec_{\timeValue+1} = \thetaVec_\timeValue + \frac{1}{\timeValue + 1}(\nonLinFunction(\thetaVec_\timeValue) - \thetaVec_\timeValue + M_\timeValue)	
\end{equation}
where, $F(\cdot)$ is a contraction map~(for sufficiently large $\deltaboth$) and $M_\timeValue$ is a martingale difference. Stochastic averaging theory~\cite[Chapter~5]{kushner2003} tells us that the non-linear stochastic dynamical system~\eqref{eq:theta_stochastic_approximation} can be approximated by a deterministic ordinary differential equation~(ODE) $\dot{\thetaVec} = \nonLinFunction(\theta) - \thetaVec$. Since~$\nonLinFunction(\cdot)$ is a contraction map, the ODE $\dot{\thetaVec} = \nonLinFunction(\theta) - \thetaVec$ has a globally asymptotically stable equilibrium~$\thetaVec_{*}$ which is the unique fixed point of the function~$\nonLinFunction(\cdot)$. This implies that the sequence $\thetaVec_\timeValue, \timeValue = 1, 2, \dots$ converges almost surely~(i.e.,~with probability~$1$) to the globally asymptotically stable equilibrium~$\thetaVec_{*}$ irrespective of the initial state~$\thetaVec_0$.\footnote{Note that the evolution of $\thetaVec_\timeValue$ in~(\ref{eq:theta_stochastic_approximation}) is of the same form as the well-known Robbins-Monroe algorithm applied to finding the fixed point of the function~$\nonLinFunction(\cdot)$. This provides further context for understanding the intuition behind the proof of Theorem~\ref{th:convergence_DMPA}. } The proof of part~ii of the theorem is also based on expressing the evolution of $\frac{\numNodesIndegreeTime(\calRed_\timeValue)}{\timeValue}$, $\frac{\numNodesOutdegreeTime(\calRed_\timeValue)}{\timeValue}$ as stochastic dynamical systems and, draws on the part~i. Further, the assumptions in Theorem~1~($\deltaIn = \deltaOut$ and $\matrixAnyEvent_{1} = \matrixAnyEvent_{2} = \matrixAnyEvent_{3}$) are imposed only to keep the notation manageable and can be easily relaxed without affecting the essence of Theorem~\ref{th:convergence_DMPA}. 
	
While the proof of Theorem~\ref{th:convergence_DMPA} is inspired by the ideas in~\cite{ bollobas2003directed, avin2015homophily, avin2020mixed}, there are several key differences. First, we are dealing with a directed, bi-populated graph which is different from the models considered in \cite{avin2015homophily, avin2020mixed}~(undirected graph with no network densification) and \cite{bollobas2003directed}~(directed graph with a homogeneous population). Further, we establish the almost sure convergence~(i.e.,~convergence with probability~$1$) of the quantities considered in Theorem~\ref{th:convergence_DMPA} compared to the convergence in expectation established in~\cite{avin2015homophily, avin2020mixed, bollobas2003directed}.

\vspace{0.2cm}
\noindent
{\bf Calculating the asymptotic power-law exponent values recursively:} Part~i of Theorem~\ref{th:convergence_DMPA} states that the normalized sum of in-degrees~$\thetaIn_\timeValue$ and the normalized sum of out-degrees~$\thetaOut_\timeValue$  of red nodes (defined in (\ref{eq:thetaTime})) converge~(for sufficiently large $\deltaboth$) to unique values $\thetaIn_{*}, \thetaOut_{*}$ respectively, with probability~$1$. Part~ii of Theorem~\ref{th:convergence_DMPA} implies that the in- and out- degree distributions of the red group are almost surely power-laws. The closed-form expressions for power-law exponents of the red group depend only on the model parameters $\redBirthProb, \probEventOne , \probEventTwo, \deltaboth, E$ and the unique value $\thetaVec_{*}$ to which the sequence $\thetaVec_\timeValue, \timeValue = 1,2,\dots$ converges. Analytically finding the unique value $\thetaVec_{*} = [\thetaIn_{*}, \thetaOut_{*}]^T$ involves solving the non-linear system of equations $\nonLinFunction(\thetaVec) = \thetaVec$ which is difficult except in special cases~(some of which are discussed in Sec.~\ref{sec:properties}). However, since the contraction map~$\nonLinFunction(\cdot)$ is known in closed form~(see Appendix~\ref{appendix:proof_convergence_DMPA}), finding the fixed point $\thetaVec_{*} = [\thetaIn_{*}, \thetaOut_{*}]^T$ can be easily achieved in an alternative way - the recursion $\thetaVec_{k+1} = \nonLinFunction(\thetaVec_k),\, k = 1,2,\dots$ (with arbitrary initial state $\thetaVec_0 \in [0,1]^2$) converges to the unique fixed point of $\nonLinFunction(\cdot)$ by the contraction mapping~(Banach fixed point) theorem. Hence, for any set of values for the parameters~($\redBirthProb, \probEventOne , \probEventTwo, \matrixAnyEvent, \deltaboth$) of the DMPA model, the unique value $\thetaVec_{*} = [\thetaIn_{*}, \thetaOut_{*}]^T$ stated in part~i of the Theorem~\ref{th:convergence_DMPA} can be easily found via a recursion. Then, substituting $\thetaVec_{*} = [\thetaIn_{*}, \thetaOut_{*}]^T$ and the values of parameters~$\redBirthProb, \probEventOne , \probEventTwo, \matrixAnyEvent, \deltaboth$ in (\ref{eq:power_law_form}) yields the power-law exponents of the in- and out- degree distributions of the red group. Since the colors (blue, red) are assigned without loss of generality, the power-law degree distributions of the blue group~($\numNodesIndegreeLimit(\calBlue), \numNodesOutdegreeLimit(\calBlue)$) also can be found by substituting ${\thetaIn_{*} \leftarrow (1-\thetaIn_{*})}$, ${\thetaOut_{*} \leftarrow (1-\thetaOut_{*})}$, ${\redBirthProb \leftarrow (1-\redBirthProb)}$, ${\homophilyBlueAnyEvent \leftarrow \homophilyRedAnyEvent}$, ${\homophilyRedAnyEvent \leftarrow \homophilyBlueAnyEvent}$ in (\ref{eq:power_law_form}). Therefore, we are able to calculate the power-law exponents of the in- and out- degree distributions of both red and blue groups for any parameter configuration of the DMPA model.

\section{Insight  into the Interplay between Structure and Dynamics}
\label{sec:properties}
So far, we have established~(in Sec.~\ref{sec:convergence}) that the asymptotic in- and out- degree distributions of each group~(blue, red) follow power-laws under the DMPA model and discussed how the exact values of the power-law exponents can be calculated using an iterative procedure. In this context, the aim of this section is to discuss the insights that can be obtained via the calculated power-law exponent values and how they can be useful for modeling the glass ceiling effect. We will first explore the quantities given in Theorem~\ref{th:convergence_DMPA}~(i.e.,~$\thetaVec_{*} = [\thetaIn_{*}, \thetaOut_{*}]^T$ and the power-law exponents) for several special cases that lead to simpler, intuitive analytical expressions. Then, other cases are discussed using values of $\thetaVec_{*} = [\thetaIn_{*}, \thetaOut_{*}]^T$ and the power-law exponents that are calculated using the recursive method stated in Sec.~\ref{sec:convergence}. Finally, we will use the obtained results to characterize the conditions under which the glass ceiling effect emerge in directed networks. 

Let us first consider the special cases where the power-law exponents given in Theorem~\ref{th:convergence_DMPA} simplify to more insightful expressions. 
\begin{theorem}[Power-law exponents of the degree distributions in special cases]
	\label{th:exponents_for_simplified_cases}
	Consider the DMPA model $\mathrm{DMPA}(\redBirthProb, \probEventOne , \probEventTwo, \matrixEventOne , \matrixEventTwo , \matrixEventThree, \deltaIn, \deltaOut)$ and assume that $\deltaIn = \deltaOut = \deltaboth$ and,
	\begin{equation}
	\matrixAnyEvent_{1} = \matrixAnyEvent_{2} = \matrixAnyEvent_{3} = \matrixAnyEvent = \begin{bmatrix}
	\homophilyBlueAnyEvent & 1 - \homophilyRedAnyEvent \\
	1 - \homophilyBlueAnyEvent & \homophilyRedAnyEvent \end{bmatrix}.
	\end{equation}
	Further, assume that $\deltaboth > \deltaboth^{*}$ where $\deltaboth^{*}>0$ is the constant guaranteed to exist by Theorem~\ref{th:convergence_DMPA}. Then, the following statements hold:
	\begin{compactenum}[(i.)]
		\item Let $\homophilyBlueAnyEvent = \homophilyRedAnyEvent = 0.5$~(i.e.,~both red and blue groups are unbiased). Then,
		\begin{equation}
		\label{eq:both_unbiased}
		\thetaIn_{*} = \thetaOut_{*} = \redBirthProb,\quad \exponentInConstant(\calRed) = 	\exponentInConstant(\calBlue) = \frac{1-\probEventTwo}{\deltaboth(\probEventOne +\probEventTwo ) + 1}, \quad 	\exponentOutConstant(\calRed) = 	\exponentOutConstant(\calBlue) = \frac{1-\probEventOne }{\deltaboth(\probEventOne +\probEventTwo ) + 1}
		\end{equation}
		
		\item Let $\homophilyBlueAnyEvent = \homophilyRedAnyEvent = 1$~(i.e.,~both groups are fully homophilic) and $\probEventTwo  = 1-\probEventOne $~(i.e.,~no edges are added between existing edges). Then,
		\begin{align}
		\label{eq:both_fully_homophilic}		
		\thetaIn_{*} = \thetaOut_{*} = \redBirthProb,\quad \exponentInConstant(\calRed) = 	\exponentInConstant(\calBlue) = \frac{\probEventOne }{\deltaboth + 1}, \quad 	\exponentOutConstant(\calRed) = 	\exponentOutConstant(\calBlue) = \frac{1-\probEventOne }{\deltaboth + 1}
		\end{align}
		
		\item Let $\homophilyBlueAnyEvent = \homophilyRedAnyEvent = 0$~(i.e.,~both groups are fully heterophilic) and $\probEventTwo  = 1-\probEventOne $~(i.e.,~no edges are added between existing edges). Then,
		\begin{equation}
			\label{eq:both_fully_heterophilic}	
			\begin{aligned}
				&\thetaIn_{*} = \redBirthProb(1-2\probEventOne) + \probEventOne,    &&\exponentInConstant(\calRed) =\frac{\probEventOne (1- \redBirthProb)}{\redBirthProb(1+\deltaboth) + \probEventOne (1-2\redBirthProb)},	&&\exponentOutConstant(\calRed) = \frac{(1-\probEventOne )(1-\redBirthProb)}{\redBirthProb(\deltaboth-1) + \probEventOne (2\redBirthProb-1)+1}	\\
				&\thetaOut_{*} = 1-\thetaIn_{*} , &&\exponentInConstant(\calBlue) = \frac{\probEventOne  \redBirthProb}{(1-\redBirthProb)(1+\deltaboth) + \probEventOne (2\redBirthProb-1)}, 	&&\exponentOutConstant(\calBlue) = \frac{(1-\probEventOne )\redBirthProb}{(1-\redBirthProb)(\deltaboth-1) + \probEventOne (1-2\redBirthProb)+1}
			\end{aligned}
		\end{equation}

		\item Let $\homophilyBlueAnyEvent = 0$~(i.e.,~blue group is fully heterophilic) and $\homophilyRedAnyEvent = 1$~(i.e.,~red group is fully homophilic). Then,
		\begin{equation}
		\label{eq:homophilic_and_heterophilic}		      
			\begin{aligned}
			&\thetaIn_{*} = \frac{\redBirthProb(\probEventOne\deltaboth(1- (\probEventOne+\probEventTwo)(1-\redBirthProb) )+ \probEventTwo(1+\deltaboth))}{\probEventOne(1-\redBirthProb + \deltaboth) + \probEventTwo(1+\deltaboth)},  &&\exponentInConstant(\calRed) = \frac{1-\probEventOne(1-\redBirthProb) -\probEventTwo}{\deltaboth(\probEventOne+\probEventTwo) + 1},	&&\exponentOutConstant(\calRed) = 	 \frac{1-\probEventOne}{\probEventOne(\redBirthProb(1+\deltaboth) -1)  + \probEventTwo\redBirthProb\deltaboth
				 + 1}\\
			&\thetaOut_{*} = 1-\probEventOne (1-\redBirthProb) , &&\exponentInConstant(\calBlue) = \frac{1-\probEventOne(1-\redBirthProb) -\probEventTwo}{\deltaboth(\probEventOne+\probEventTwo) + 1},	&&\exponentOutConstant(\calBlue) = 0
			\end{aligned}
		\end{equation}	
		
	\end{compactenum}
\end{theorem}

\begin{figure}
	\centering
	\includegraphics[width=\linewidth]{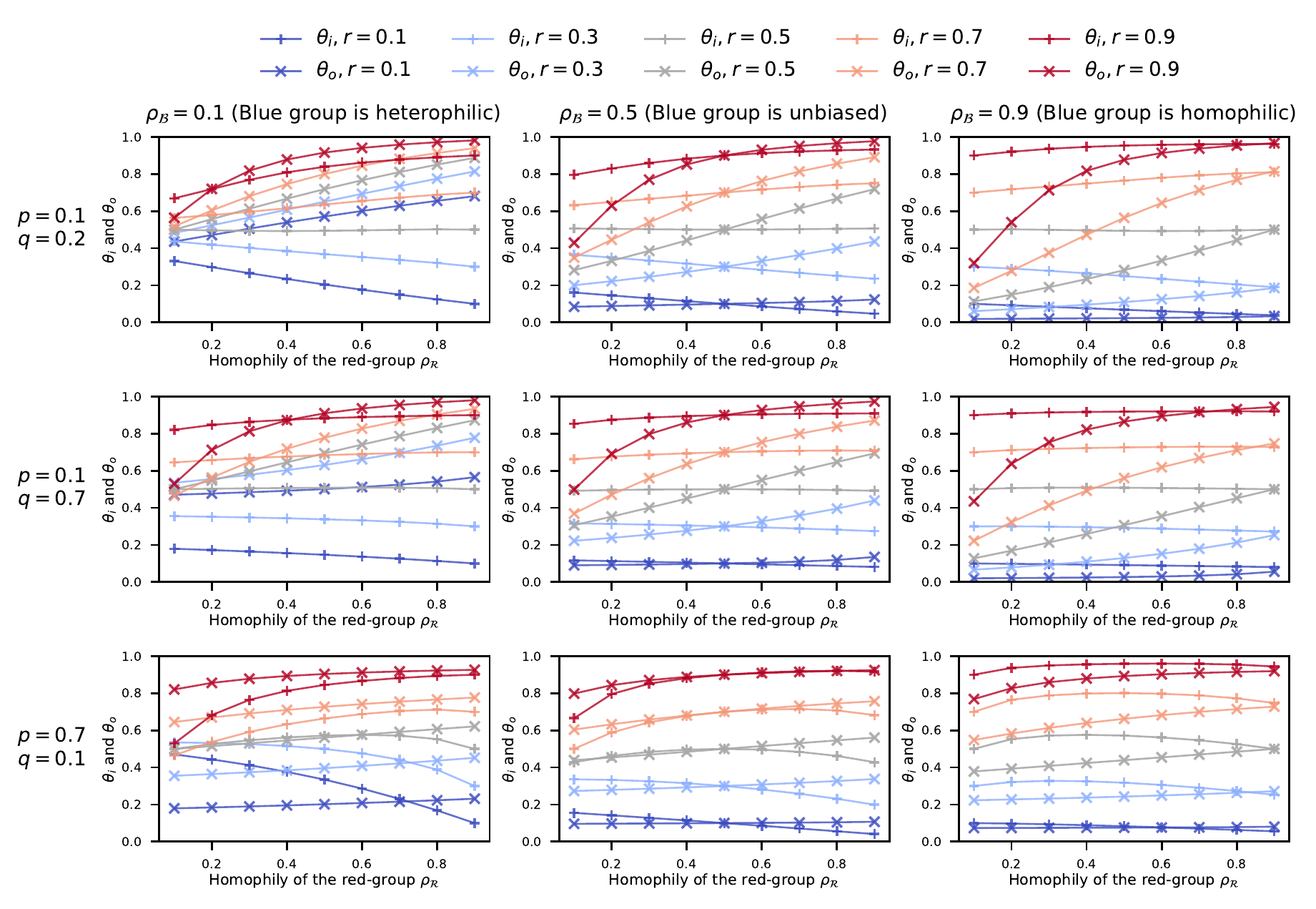}
	\caption{Figure shows variation of $\thetaIn_{*}$~(the unique asymptotic value of the fraction of the in-degrees of red nodes~$\thetaIn_\timeValue$ defined in~(\ref{eq:thetaTime})) and $\thetaOut_{*}$~(the unique asymptotic value of the fraction of the out-degrees of red nodes~$\thetaOut_\timeValue$ defined in~(\ref{eq:thetaTime})) with the homophily of the red-group~$\homophilyRedAnyEvent$. The three columns correspond to three different values of the homophily of the blue groups: $\homophilyBlueAnyEvent = 0.1$~(blue group is \emph{heterophilic} - prefers to follow red-individuals), $\homophilyBlueAnyEvent = 0.5$~(blue group is \emph{unbiased} - prefers to follow both red and blue individuals equally), $\homophilyBlueAnyEvent = 0.9$~(blue group is \emph{homophilic} - prefers to follow blue-individuals). The three rows correspond to various values of $\probEventOne$ and $\probEventTwo$ specify the probabilities of the three events that constitute the growth of the network~(illustrated in Fig.~\ref{fig:model_diagram}). In each subplot, the different colors correspond to different values of the parameter $\redBirthProb$ which determine the asymptotic fraction of red nodes in the network. The values of  $\thetaIn_{*}, \thetaOut_{*}$ are calculated using the procedure outlined in Sec.~\ref{sec:convergence}~(i.e.,~iteratively using the recursion $\thetaVec_{\timeValue+1} = \nonLinFunction(\thetaVec_\timeValue)$ where $\nonLinFunction(\cdot)$ is the contraction map in (\ref{eq:theta_stochastic_approximation})).}
	\label{fig:theta}
\end{figure}
\begin{figure}
	\centering
	\includegraphics[width=\linewidth]{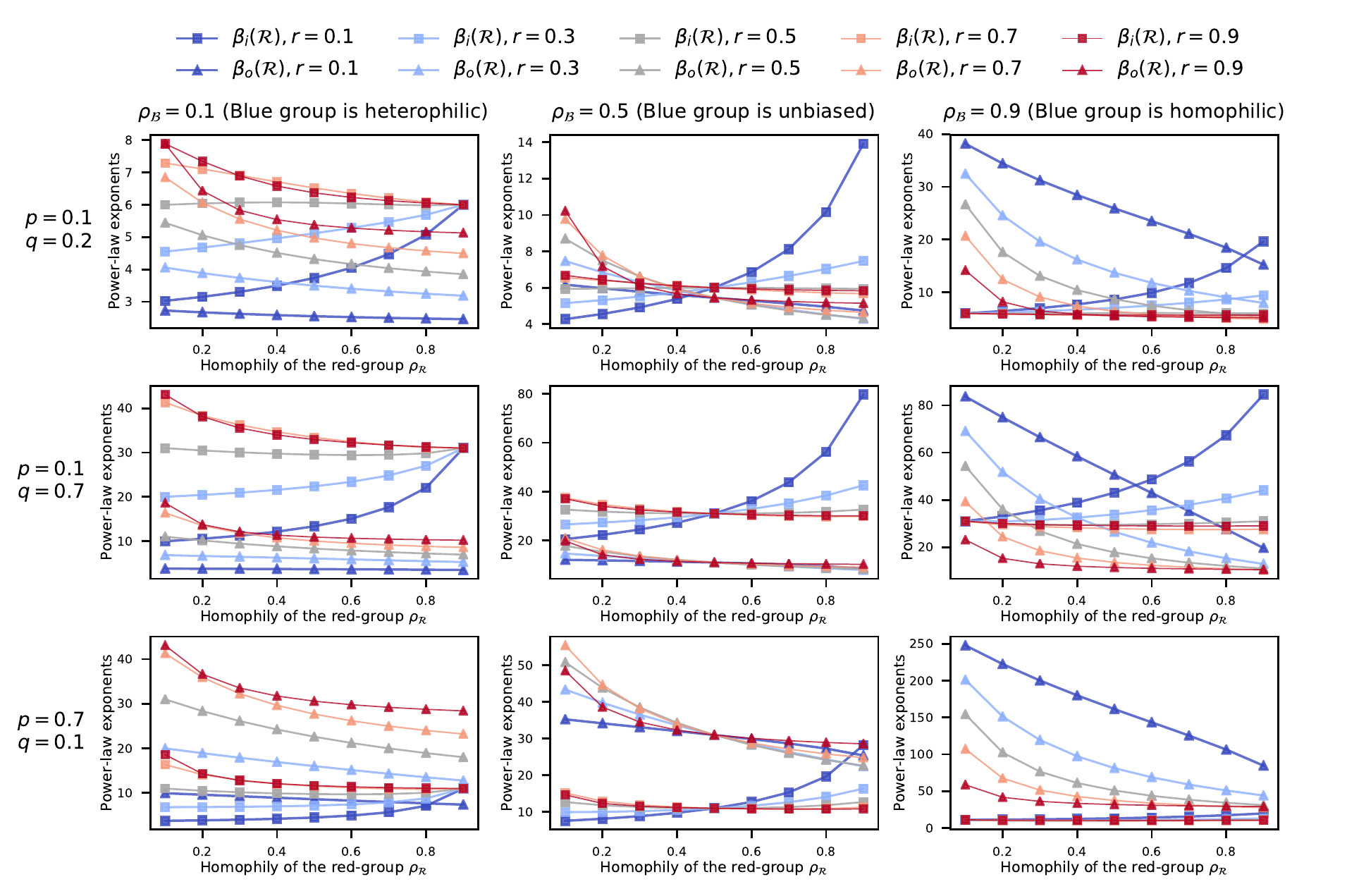}
	\caption{Figure shows variation of the power-law exponents of the asymptotic in-degree distribution~($1+\frac{1}{\exponentInConstant(\calRed)}$) and the out-degree distribution~~($1+\frac{1}{\exponentInConstant}(\calRed)$) of the red-group for various parameter values of the DMPA model. Similar to Fig.~\ref{fig:theta}, the three columns correspond to different values of the homophily of the blue group~$\homophilyBlueAnyEvent$ and the three rows correspond to three different pairs of values of $\probEventOne, \probEventTwo$~(that specify the probabilities of the events that constitute the growth of the network). The power-law exponent values are calculated by substituting the values of $\thetaIn_{*}, \thetaOut_{*}$ shown in Fig.~\ref{fig:theta} and the parameter values in the expressions given in (\ref{eq:exponentInContant}) and (\ref{eq:exponentOutContant}). The plots complement the Theorem~\ref{th:exponents_for_simplified_cases} and the insights obtained from the figure and Theorem~\ref{th:exponents_for_simplified_cases} are discussed in the context of each other in Sec.~\ref{subsec:discussion}.}
	\label{fig:pl_coefficients}
\end{figure}

The results given in Theorem~\ref{th:exponents_for_simplified_cases} are for special cases where each group~(red, blue) is either unbiased, fully heterophilic or, fully homophilic~i.e.,~$\homophilyBlueAnyEvent, \homophilyRedAnyEvent \in \{0,0.5,1\}$. Fig.~\ref{fig:theta} shows the variation of $\thetaIn_{*}, \thetaOut_{*}$~(stated in part~i of Theorem~\ref{th:convergence_DMPA}) with homophily of the red-group~$\homophilyRedAnyEvent$ for several other scenarios~(that are not covered in Theorem~\ref{th:exponents_for_simplified_cases}). Similarly, Fig.~\ref{fig:pl_coefficients} shows the variation of the power-law exponent of the in-degree distribution~($1+\frac{1}{\exponentInConstant(\calRed)}$) and the out-degree distribution ~($1+\frac{1}{\exponentOutConstant(\calRed)}$) of the red-group~(given in part~ii of Theorem~\ref{th:convergence_DMPA}) with the homophily of the red-group~$\homophilyRedAnyEvent$. In both Fig.~\ref{fig:theta} and Fig.~\ref{fig:pl_coefficients}, the three columns correspond to three different values of the homophily of the blue groups: $\homophilyBlueAnyEvent = 0.1$~(blue group is \emph{heterophilic} - prefers to follow red-individuals), $\homophilyBlueAnyEvent = 0.5$~(blue group is \emph{unbiased} - prefers to follow both red and blue individuals equally), $\homophilyBlueAnyEvent = 0.9$~(blue group is \emph{homophilic} - prefers to follow blue-individuals). Further, the three rows of both Fig.~\ref{fig:theta} and Fig.~\ref{fig:pl_coefficients} correspond to various values of $\probEventOne , \probEventTwo$ which characterize the growth dynamics of the directed graph. The first row~($\probEventOne <\probEventTwo < 1-\probEventOne -\probEventTwo$) corresponds to the case where existing nodes following each other is the most likely event with the second one being the new nodes following existing nodes~e.g.~Twitter, author-citation networks, etc. The second row~($\probEventOne <1-\probEventOne -\probEventTwo < \probEventTwo$) corresponds to the case where new nodes following existing nodes is the most likely event with the second one being the existing nodes following each other~e.g.~a new research field where new authors are joining frequently and follow existing authors, a fast growing new social network. The third row~($\probEventOne > 1-\probEventOne -\probEventTwo > \probEventTwo$) is the case where new nodes joining and being followed by existing nodes is the most likely event with the second one being the existing nodes following each others.  The values shown in both Fig.~\ref{fig:theta} and Fig.~\ref{fig:pl_coefficients} are calculated using the recursive method detailed in Sec.~\ref{sec:convergence}. Several insights can be drawn from Fig.~\ref{fig:theta}, Fig.~\ref{fig:pl_coefficients} and Theorem~\ref{th:exponents_for_simplified_cases}.\footnote{  
Some of the values in Fig.~\ref{fig:pl_coefficients} fall outside the range of power-law exponents that are typically observed in nature~(i.e.,~the interval $[2,3.5]$). This is because the parameter regimes illustrated in Fig.~\ref{fig:pl_coefficients} were chosen to illustrate the patterns observed in all possible cases~(some of which may not be prevalent in nature such as purely homophilic or heterophilic behaviors, negligibly smaller minority group sizes etc.); when the power-law exponents are larger, they could be interpreted as thin tailed distributions. The top-left sub-figure (which corresponds to parameter configurations observed in many real-world networks~\cite{nettasinghe2020emergence}) has power-law exponent values that align mostly with the ranges observed in real-world networks.}

\subsection{Discussion of Theorem~\ref{th:exponents_for_simplified_cases} and Fig.~\ref{fig:pl_coefficients}}
\label{subsec:discussion}
In this subsection, we will discuss the Theorem~\ref{th:exponents_for_simplified_cases} and Fig.~\ref{fig:pl_coefficients} in order to understand the insights that they provide. 

Let us first consider the case where both groups are unbiased~(case~i in Theorem~\ref{th:exponents_for_simplified_cases}). Eq.~\eqref{eq:both_unbiased} shows that the asymptotic fraction of the total in-degree~($\thetaIn_{*}$) and the  asymptotic fraction of the total out-degree~($\thetaOut_{*}$) of the red group is both equal to $\redBirthProb$~(probability that a new node is given the color red). Hence, only the group size determines the fraction of in-degrees and out-degrees owned by each group in case~i. 
However, the two groups are equal to each other in terms of the degree distributions~i.e.,~in-degree distribution~(resp.~out-degree distribution) of the red group has the same power-law exponent as the in-degree distribution~(resp.~out-degree distribution) of the blue group. Further, the power-law exponents of both in- and out- degree distributions do not depend on $\redBirthProb$ in case~i. This implies that when both groups are  unbiased, in- and out- degree distribution of each group is not affected by the size of groups.  Also, in case~i, the out-degree distribution of each group has a heavier tail than the in-degree distribution of that group if and only if $\probEventOne  < \probEventTwo$.\footnote{Due to the power-law form of the distributions, $\probEventOne  < \probEventTwo$ in case~i of Theorem~\ref{th:exponents_for_simplified_cases} is equivalent to saying that the out-degree distribution dominates the in-degree distribution in the \emph{monotone likelihood ratio order} sense~i.e.,~the ratio of the probability mass function of the out-degree distribution to the probability mass function of the in-degree distribution increases in the degree~(for both red and blue groups). } Finally, in case~i, decreasing the value $\probEventOne + \probEventTwo$~(i.e.,~the probability of the first two events where a new node is added) leads to larger power-law exponents and hence, thinner tails. This observation suggests that preferential-attachment among existing nodes makes the tails heavier compared to preferential-attachment between existing nodes and new nodes~i.e.,~adding edges more frequently between existing nodes of the graph~(i.e.,~the third event happening more frequently) leads to heavier tails in both groups. In fact, this can be observed from Fig.~\ref{fig:pl_coefficients} as well - the power-law coefficients in the first row~(where $\probEventOne + \probEventTwo = 0.3$) are smaller~(i.e.,~tails are heavier) compared to the second and third row~(where $\probEventOne + \probEventTwo = 0.7$). 

Let us now look at the case where both groups are fully homophilic and the third event~(edges being added between existing nodes) has a zero probability~(case~ii in Theorem~\ref{th:exponents_for_simplified_cases}). Eq.~\eqref{eq:both_fully_homophilic} shows that, similar to case~i,  the fraction of the total in- and out- degrees that belong to red-group are both equal to the asymptotic fraction of red nodes in the network~(i.e.,~$\thetaIn_{*} = \thetaOut_{*} = \redBirthProb$). Further, the two groups~(red, blue) are also equal to each other in terms of the degree distributions similar to case~i. If $\probEventOne  < 0.5$, then the  out-degree distribution each group has a heavier tail than the in-degree distribution of that group~(i.e.,~the out-degree distribution dominates the in-degree distribution in a monotone likelihood ratio order sense). This observation confirms that increasing $\probEventOne$ makes the tail of the out-degree distribution heavier and the tail of the in-degree distribution thinner as we saw from case~i as well - this observation is common for all four cases in Theorem~\ref{th:exponents_for_simplified_cases}. The reason is intuitive: when $\probEventOne$ is increased, new nodes with an in-degree of $1$ are frequently added to the graph, causing the bulk of the in-degree distributions to be concentrated around smaller degrees while at the same causing bulk of the out-degree distributions to be moved to larger degrees~(since new nodes will be following existing nodes).

\begin{figure*}
	\begin{subfigure}[b]{0.45\textwidth}
		\centering
		\includegraphics[trim=0.5in 0.35in 0.1in 0.4in,clip,width=\textwidth]{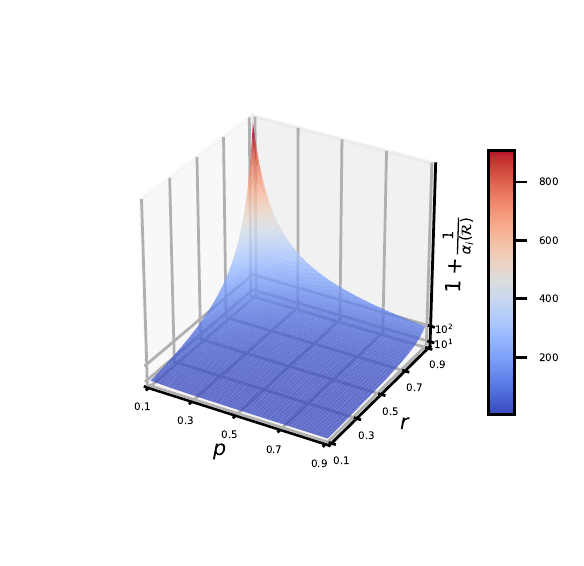}
		\caption[]%
		{Power-law exponent of in-degree distribution}
	\end{subfigure} \hspace{0.2in}
	\begin{subfigure}[b]{0.45\textwidth}
		\centering
		\includegraphics[trim=0.5in 0.35in 0.1in 0.4in,clip,width=\textwidth]{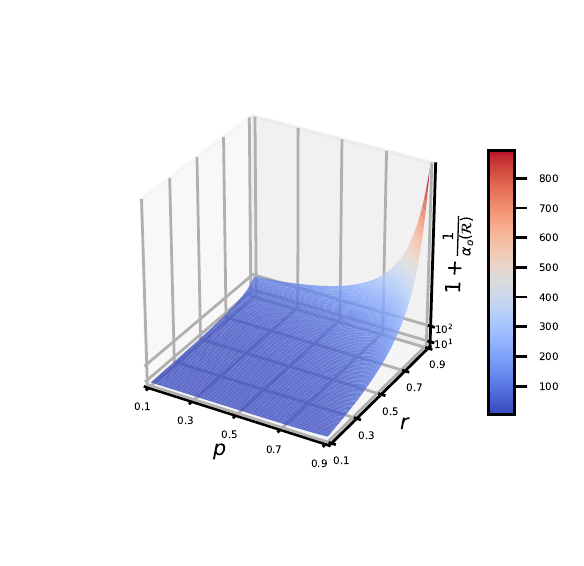}
		\caption[]%
		{Power-law exponent of out-degree distribution}
	\end{subfigure}
	\caption{
		Power-law exponents of the in-degree distribution~($1+\frac{1}{\exponentInConstant(\calRed)}$) and the out-degree distribution~($1+\frac{1}{\exponentOutConstant(\calRed)}$) of the red-group when both red and blue groups are heterophilic and the third event~(adding edges between existing nodes) has zero probability~(case~iii in Theorem~\ref{th:exponents_for_simplified_cases}). The plots show that, when both groups are heterophilic,  both in- and out- degree distributions have heavier tails when the size of the group is smaller. Further, increasing the value of $\probEventOne$~(probability of the first event) makes the tail of the in-degree distribution heavier~(i.e.,~smaller exponent) and the tail of the out-degree distribution lighter~(i.e.,~larger exponent). 
	} 
	\label{fig:case_three}
\end{figure*}
Let us now turn to the case where both groups are fully heterophilic and the third event~(edges being added between existing nodes) has a zero probability~(case~iii in Theorem~\ref{th:exponents_for_simplified_cases}). Case~iii differs from the first two cases in several ways: $\thetaIn_{*}, \thetaOut_{*}$ depend on both $\probEventOne, \redBirthProb$ and, the two groups are not equivalent to each other in terms of in- and out- degree distributions. This is because the group size was not important in previous cases since either the node colors were ignored during link formation~(case~i) or nodes only followed others who were within their own group~(case~ii). However, in case~iii, nodes follow other nodes from the opposite group only and therefore, the group sizes affect the degree distributions of each group. Further, when $\redBirthProb = 0.5$, the power-law exponents for case~iii become the same as those in case~ii~(i.e.,~$\exponentInConstant(\calRed) = 	\exponentInConstant(\calBlue) = \frac{\probEventOne }{\deltaboth + 1}, \exponentOutConstant(\calRed) = 	\exponentOutConstant(\calBlue) = \frac{1-\probEventOne }{\deltaboth + 1}$) since the group sizes are equal. Interestingly, we can see that when the asymptotic size of the red group~$\redBirthProb$ becomes smaller, $\exponentOutConstant(\calRed)$ gets larger - this implies that the tail of the out-degree distribution of a group becomes heavier as the group size become smaller. The in-degree distribution shows a similar trend as well. This can be observed from Fig.~\ref{fig:case_three} which illustrates the variation of the power-law exponents by using the expressions in (\ref{eq:both_fully_heterophilic}).

\begin{figure*}
	\begin{subfigure}[b]{0.45\textwidth}
		\centering
		\includegraphics[trim=0.5in 0.35in 0.1in 0.4in,clip,width=\textwidth]{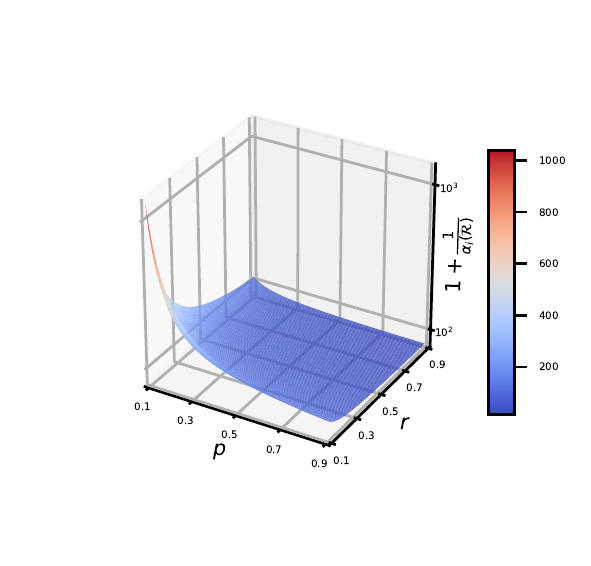}
		\caption[]%
		{Power-law exponent of in-degree distribution}
	\end{subfigure} \hspace{0.2in}
	\begin{subfigure}[b]{0.45\textwidth}
		\centering
		\includegraphics[trim=0.5in 0.35in 0.1in 0.4in,clip,width=\textwidth]{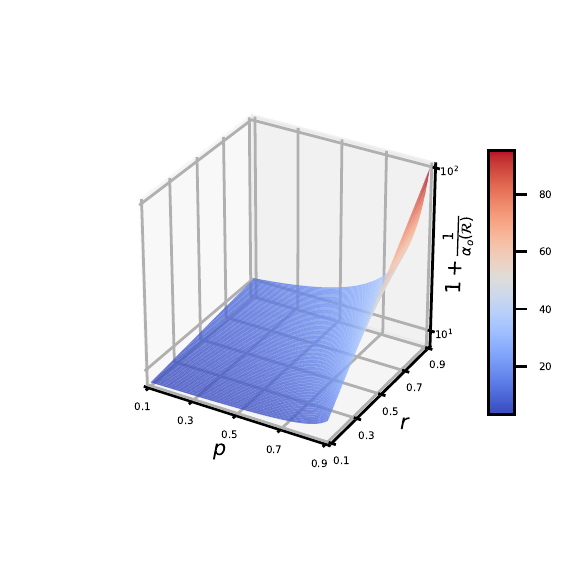}
		\caption[]%
		{Power-law exponent of out-degree distribution}
	\end{subfigure}
	\caption{
		Power-law exponents of the in-degree distribution~($1+\frac{1}{\exponentInConstant(\calRed)}$) and the out-degree distribution~($1+\frac{1}{\exponentOutConstant(\calRed)}$) of the red-group when the red group is homophilic~($\homophilyRedAnyEvent = 1$), the blue group is heterophilic~($\homophilyBlueAnyEvent = 0$) and the third event~(adding edges between existing nodes) has zero probability~(case~iv in Theorem~\ref{th:exponents_for_simplified_cases} with $\probEventTwo = 1-\probEventOne$). The plots show that, when red group is homophilic and the blue group is heterophilic, in degree distribution of the red group becomes heavier with size of the red-group $\redBirthProb$ whereas the out-degree distribution of the red group becomes lighter with the size of the red-group~$\redBirthProb$.
	} 
	\label{fig:case_four}
\end{figure*}

Let us now consider the final case~(case~iv in Theorem~\ref{th:exponents_for_simplified_cases}) where the red-group is fully homophilic and the blue group is fully heterophilic. In case~iv, $\exponentOutConstant(\calBlue) =0$, implying that all blue nodes have zero out-degree (with probability one) in this setting. Further, (\ref{eq:homophilic_and_heterophilic}) shows that when the red-group becomes smaller~(i.e.,~$\redBirthProb$ decreases), the in-degree distribution of the red-group becomes lighter while the out-degree distribution of the red-group becomes heavier. Hence, when $\redBirthProb$ is smaller, the smaller red group will consist of high out-degree~(i.e.,~followed by many others) and low in-degree~(i.e.,~follow few others) nodes; as we will see in the next subsection, this leads to a glass ceiling effect for the blue group. This can also be observed from Fig.~\ref{fig:case_four} where we also set $\probEventTwo = 1-\probEventOne$. This observation is in contrast to the case~iii~(both groups were heterophilic) where both in-degree and out-degree distributions of the red-group became lighter with size of the red-group~$\redBirthProb$.

\subsection{The Glass Ceiling Effect in Directed Networks: An illustration of the practical use of the DMPA model and its analysis}
\label{subsec:GCE}
So far, we discussed the DMPA model, derived the power-law exponents of the in- and out- degree distribution of each group in the DMPA model and, discussed the insight that the results provide.  In this context, the aim of this subsection is to analyze the glass ceiling effect in the context of directed graphs utilizing the results derived earlier. In general, our aim is to highlight how our results are useful for understanding the sociological consequences that have been limited so far to the context of undirected networks.

The (Tail) Glass ceiling effect in undirected graphs has been defined in \cite[Definition~2]{avin2015homophily} as follows: An undirected graph sequence $\{\graph(\timeValue)\}_{t \in \mathbb{N}}$ exhibits a glass ceiling effect if there exists an increasing function $k(\timeValue)$ such that $\lim\limits_{\timeValue \rightarrow \infty} \frac{\mathbb{E}\left\{ \mathrm{top}_{k(\timeValue)}\left(\calRed\right)\right\}}{\mathbb{E}\left\{ \mathrm{top}_{k(\timeValue)}\left(\calBlue\right)\right\}}$ where $\mathrm{top}_{k(\timeValue)}\left(\calRed\right)$, $\mathrm{top}_{k(\timeValue)}\left(\calBlue\right)$ denotes the number nodes with degree of at least $k$ in red and blue groups, respectively. However, in directed graphs, we need need to consider both the in- and out- degrees when defining the glass ceiling effect. For this, we adopt the following modified definition. 

\begin{definition}{{\normalfont (Glass Ceiling Effect in Directed Graphs)}}
	\label{def:DGCE}
	A directed graph sequence $\{\graph_\timeValue\}_{t \in \mathbb{N}}$ exhibits a glass ceiling effect towards the red group~$\calRed_\timeValue$ if,
	 \begin{equation}
	\lim\limits_{k\rightarrow \infty} \lim\limits_{\timeValue \rightarrow \infty} \left(\frac{ \mathrm{top}^{(o)}_{k}\left(\calRed_\timeValue\right)}{ \mathrm{top}^{(i)}_{k}\left(\calRed_\timeValue\right)} \times \frac{ \mathrm{top}^{(i)}_{k}\left(\calBlue_\timeValue\right)}{ \mathrm{top}^{(o)}_{k}\left(\calBlue_\timeValue\right)}\right) = 0
	\end{equation} almost surely~(i.e.,~with probability~1), where $\mathrm{top}^{(o)}_{k}\left(\calRed_\timeValue\right)$, $\mathrm{top}^{(i)}_{k}\left(\calRed_\timeValue\right)$~(respectively, $\mathrm{top}^{(o)}_{k}\left(\calBlue_\timeValue\right)$, $\mathrm{top}^{(i)}_{k}\left(\calBlue_\timeValue\right)$) denote the number of red~(respectively, blue) nodes with out-degree and in-degree of at least $k$, respectively.
\end{definition}

The intuition behind the Definition~\ref{def:DGCE} is as follows. Note that $\mathrm{top}^{(o)}_{k}\left(\calRed\right)$ is the number of red nodes who are followed by at least $k$ nodes and thus, represents the set of individuals who have risen to ``fame" or ``popularity". Dividing $\mathrm{top}^{(o)}_{k}\left(\calRed\right)$ by $\mathrm{top}^{(i)}_{k}\left(\calRed\right)$ accounts for the fact that the popularity could be a result of many connections being reciprocal~(i.e.,~simply because they follow many others who also have reciprocated the relationship).\footnote{For example, in an author citation network, this accounts for the fact that survey papers accrue a large number of citations in most instances due to their lengthier bibliographies.} Thus, ${ \mathrm{top}^{(o)}_{k}\left(\calRed\right)}/{\mathrm{top}^{(i)}_{k}\left(\calRed\right)}$ can be viewed as an indication of the tail heaviness of the out-degree distribution compared to that of the in-degree distribution of the red group. A similar interpretation holds for the blue group as well. Hence, according to Definition~\ref{def:DGCE}, the red group faces a glass ceiling effect if the weight of tail of its out-degree distribution normalized by the weight of the tail of its in-degree distribution is asymptotically~(i.e.,~as $\timeValue, k \rightarrow \infty$) is negligible compared to that of the blue group~(almost surely). 

\begin{remark}[Directed edges are necessary for the emergence of the glass ceiling effect]
Note that the Definition~\ref{def:DGCE} is specific to directed graphs in the sense that the glass ceiling effect does not arise in undirected graphs according to Definition~\ref{def:DGCE}. In particular, the limit in the Definition~\ref{def:DGCE} will always be equal to~$1$ for any non-trivial sequence of undirected graphs. Thus, the asymmetry in edges is a necessary condition for the emergence of the glass ceiling effect in the sense of Definition~\ref{def:DGCE}.
\end{remark}

\vspace{0.2cm}
\noindent
{\bf Glass-ceiling effect under the DMPA model:} Let us now examine the Definition~\ref{def:DGCE} in the context of the DMPA model. Since we know that the asymptotic in- and out- degree distributions of the blue and red groups are power-laws almost surely~(from Theorem~\ref{th:convergence_DMPA}), we get,
\begin{align}
\lim\limits_{\timeValue \rightarrow \infty}  \left(\frac{ \mathrm{top}^{(o)}_{k}\left(\calRed_\timeValue\right)}{ \mathrm{top}^{(i)}_{k}\left(\calRed_\timeValue\right)}\right) = \frac{\numNodesOutdegreeLimit(\calRed)}{\numNodesIndegreeLimit(\calRed)}  \propto k^{\frac{1}{\exponentInConstant(\calRed)} - \frac{1}{\exponentOutConstant(\calRed)}}\,, \quad \lim\limits_{\timeValue \rightarrow \infty}  \left(\frac{ \mathrm{top}^{(o)}_{k}\left(\calBlue_\timeValue\right)}{ \mathrm{top}^{(i)}_{k}\left(\calBlue_\timeValue\right)}\right) = \frac{\numNodesOutdegreeLimit(\calBlue)}{\numNodesIndegreeLimit(\calBlue)}  \propto k^{\frac{1}{\exponentInConstant(\calBlue)} - \frac{1}{\exponentOutConstant(\calBlue)}} 
\end{align}
almost surely. An immediate consequence is the following corollary which characterizes the parameter regimes that lead to the emergence of the glass ceiling effect under the DMPA model.

\begin{corollary}[Necessary and Sufficient Conditions for the Emergence of the Glass Ceiling Effect]
	Consider the DMPA model $\mathrm{DMPA}(\redBirthProb, \probEventOne , \probEventTwo, \matrixEventOne , \matrixEventTwo , \matrixEventThree, \deltaIn, \deltaOut)$ under the same assumptions as in Theorem~\ref{th:convergence_DMPA}. Then, the red group faces a glass ceiling effect~(Definition~\ref{def:DGCE}) if and only if,
	\begin{equation}
\frac{1}{\exponentInConstant(\calRed)} - \frac{1}{\exponentOutConstant(\calRed)} <	\frac{1}{\exponentInConstant(\calBlue)} - \frac{1}{\exponentOutConstant(\calBlue)}
	\end{equation}
	where, $\exponentInConstant(\calBlue),\exponentOutConstant(\calBlue), \exponentInConstant(\calRed),\exponentOutConstant(\calRed)$ are given in Theorem~\ref{th:convergence_DMPA}.
\end{corollary}

Hence, the Theorem~\ref{th:convergence_DMPA} (and the recursive procedure discussed following it) can be used to find out if the glass ceiling effect emerge under any given a parameter configuration. In addition, Theorem~\ref{th:exponents_for_simplified_cases} and Fig.~\ref{fig:pl_coefficients} (that were discussed in detail in Sec.~\ref{subsec:discussion}) illustrate special cases which provide additional insight into the conditions that lead to the glass ceiling effect.

\section{Summary and Extensions}
Preferential-attachment~(i.e.,~preference of individuals to form links with already popular individuals) and homophily~(preference of individuals to form connections with others who have similar attributes) are two well-studied sociological consequences observed in undirected social networks. However, a model which incorporates preferential attachment and homophily into a dynamic directed network~(which evolves over time) has been absent in literature, causing the study of important sociological consequences such as the \emph{glass ceiling effect} to be limited to the setting of undirected networks. Towards this end, we proposed a dynamic model named \emph{Directed Mixed Preferential Attachment~(DMPA)} model which contains two types of nodes~(red and blue). In the DMPA model, one of the three events takes place at each time instant according to a probability distribution: a new node joins and an existing node follows the new node~(event~1), a new node joins and it follows an existing node~(event~2) and, an existing node follows another existing node~(event~3). Hence, a new link is added to the network at each time instant. The existing nodes that will potentially be included in the new link~(candidate nodes) are chosen according to preferential attachment~i.e.,~individuals that follow more people are likely to follow more and individuals that are followed by more people are likely to be followed by more. The link is added between the candidate nodes with probabilities that depend on their colors~(modeling the homophily) as well as the event that is taking place. Using tools from stochastic averaging theory, we show that this model asymptotically leads to power-law in- and out- degree distributions for both groups and derive their closed form expressions. Using the derived expressions, we explore how factors such as group size~(i.e.,~the size of the minority group compared to the majority group), homophily and the growth dynamics~(i.e.,~the probabilities assigned to the three events) affect the in- and out- degree distributions of the two groups. The results yield insight into questions such as: when does the minority group have a heavier out-degree~(or~in-degree) distribution compared to the majority?, what factors lead the out-degree distribution of a group to be heavier than its in-degree distribution?, what effect does frequent addition of new nodes with out-going~(or incoming) links have on the in- and out- degree distributions of each group? etc. Then, using the derived results, a characterization of the conditions that lead to the glass ceiling effect in directed social networks was obtained. In addition, the proposed DMPA model is amenable to computationally efficient, simulation based studies of directed graphs since the asymptotic power-law expoenents can be obtained via a simple iterative procedure~(without the need to perform computationally expensive Monte Carlo simulations that involve graphs) as a consequence of the Banach fixed point theorem.

\vspace{0.05cm}
\noindent
{\bf Extensions: } The DMPA model and its analysis that we presented open up several interesting research avenues. First, the DMPA model incorporates sociological phenomena such as preferential-attachment, homophily, group size by treating most decisions~(e.g.~color of a new node, whether or not to follow a candidate node) as Bernoulli random variables. Alternatively, these decisions could be treated as strategic~(i.e.,~rational) choices that result from maximizing a utility function whose structure represents various human behavioral traits such as risk aversion, limited attention~\cite{sims2003implications}, perception bias~\cite{alipourfard2020friendship} etc. In addition, tools such as revealed-preference allows such utility functions to be estimated by observing the choices made by individuals in growing networks and therefore, can shed light into the interplay between strategic decision making and network growth. Second, the proposed model can also be used to study how algorithmic recommendation methods can alter the growth of networks - such studies have been conducted for undirected networks~(e.g.~\cite{stoica2018algorithmic}) and may have counter-intuitive outcomes in directed networks due to the asymmetry in the links. Finally, the power-law exponents of real-world networks can be estimated using friendship paradox based estimation methods~\cite{nettasinghe2019maximum} in order to verify the predictions of the model.


\newpage 
\appendix

\section{Proof of Theorem~\ref{th:convergence_DMPA}}
\label{appendix:proof_convergence_DMPA}

\subsection{Preliminaries and the outline of the proof}
The proof of Theorem~\ref{th:convergence_DMPA} uses the following result from~\cite[Theorem~2.1]{kushner2003}.
\begin{theorem}[Convergence w.p.~1 under Martingale difference noise~\cite{kushner2003}]
	\label{th:kushner_theorem}
	Consider the algorithm
	\begin{equation}
	\label{eq:algorithm_kushner_theorem}
	x_{\timeValue + 1} = x_{\timeValue} + \epsilon_\timeValue Y_\timeValue +\epsilon_\timeValue z_\timeValue
	\end{equation}
where,  $Y_\timeValue \in \mathbb{R}^m$, $\epsilon_{\timeValue}$ denotes a decreasing step size, and $\epsilon_\timeValue z_\timeValue$ is the shortest Euclidean length needed to project $x_\timeValue$ into some compact set $H$. Assume,
	\begin{compactenum}
		\item[(C.1)]  $\sup_\timeValue |Y_\timeValue|^2 < \infty$
		
		\item[(C.2)] There exists a measurable function $\bar{g}(\cdot)$ of $x$ and random variables $\beta_\timeValue$ such that, 
		\begin{equation}
		\mathbb{E}_\timeValue Y_\timeValue = \mathbb{E}\{Y_\timeValue|x_0, Y_i, i < \timeValue\} = \bar{g}(x_\timeValue) +\beta_\timeValue
		\end{equation}
		
		\item[(C.3)] $\bar{g}(\cdot)$ is Lipschitz continuous
		
		\item[(C.4)] $\epsilon_\timeValue \geq 0, \epsilon_i \rightarrow 0$ for $i \geq 0$ and $\epsilon_{\timeValue} = 0$ for $i < 0$
		
		\item[(C.5)] $\sum_{\timeValue}\epsilon_{\timeValue}^2 < \infty, \sum_{i}\epsilon_i = \infty$
		
		\item[(C.6)] $\sum_{i}\epsilon_i\left|\beta_i\right| < \infty$	w.p.~1
	\end{compactenum}
If $\bar{x}$ is an asymptotically stable point of the ordinary differential equation~(ODE)
\begin{equation}
\label{eq:ODE_kushner_theorem}
 \dot{x} = \bar{g}(x) + z
\end{equation} where $z$ is the shortest Euclidean length needed to ensure $x$ is in $H$ and, $x_\timeValue$ is in some compact set in the domain of attraction of $\bar{x}$ infinitely often with probability $\geq \rho$, then $x_\timeValue \rightarrow \bar{x}$ with at least probability~$\rho$. 
\end{theorem} 

In the context of Theorem~\ref{th:kushner_theorem}, our approach for proving part~i of Theorem~\ref{th:convergence_DMPA} is to express the evolution of~$\thetaVec_\timeValue$ in the form of $(\ref{eq:algorithm_kushner_theorem})$ with $z_\timeValue = 0$ and, show that the assumptions C.1-C.6 are satisfied. Then, we show that the function~$\bar{g}$ is of the form $\bar{g}\left(\thetaVec_\timeValue\right) = \nonLinFunction(\thetaVec_\timeValue) - \thetaVec_\timeValue$ where $\nonLinFunction(\cdot)$ is a contraction map. Theorem~\ref{th:kushner_theorem} thus implies that there exists a globally asymptotically stable equilibrium state~(which is the unique fixed point of the contraction map~$\nonLinFunction(\cdot)$) to which the sequence $\{\thetaVec_\timeValue\}$ converges almost surely. 

The idea behind the proof of part~ii of the Theorem~\ref{th:convergence_DMPA} is also to exploit Theorem~\ref{th:kushner_theorem}. More specifically, we first express the evolution of ${\numNodesIndegreeTime\left(\calRed_{\timeValue}\right)}/{\timeValue}$~(where, ${\numNodesIndegreeTime\left(\calRed_{\timeValue}\right)}$ is the number of red-nodes with in-degree $k$ at time $\timeValue$ as defined in~(\ref{eq:numNodesDegreeTime})) in the form of $(\ref{eq:algorithm_kushner_theorem})$ with $z_\timeValue = 0$ and show that the assumptions C.1-C.6 are satisfied. Then, we show that $\numNodesIndegreeLimit(\calRed)$~(which is the equilibrium state of ${\numNodesIndegreeTime\left(\calRed_{\timeValue}\right)}/\timeValue$) satisfies a ratio property which implies that it is a power-law distribution according to Lemma~\ref{lemma:power_law_form}. The idea for obtaining the power-law form of the out-degree distribution is also similar.  
\begin{lemma}
	\label{lemma:power_law_form}
	Let $p(\cdot)$ be a discrete probability distribution such that $p(k) \propto k^{-\gamma}$~i.e.,~$p$ is a power-law distribution with the exponent $\gamma>0$. Then, $p$ satisfies,
	\begin{equation}
	\frac{p(k)}{p(k-1)} = \left(1- \frac{1}{k}\right)^{\gamma} =  1 - \frac{\gamma}{k} + O\left(\frac{1}{k^2}\right). \nonumber
	\end{equation}
\end{lemma}

\subsection{Proof of part~i of Theorem~\ref{th:convergence_DMPA}}
{\bf Notation:} We first need to establish some additional notation that we need for the proof of first part. Let,
\begin{compactitem}
		\item[] $\poneBB : $ given the event~$1$ and the new node is blue, probability that it is followed by an existing blue node
		\item[] $\poneBR: $ given the event~$1$ and the new node is blue, probability that it is followed by an existing red node
		\item[] $\poneRR : $ given the event~$1$  and the new node is red, probability that  it is followed by an existing red node
		\item[] $\poneRB : $ given the event~$1$ and the new node is red, probability that it is followed by an existing blue node.
\end{compactitem}
Analogous quantities for event~2 and event~3 are denoted by $\ptwoBB, \ptwoBR, \ptwoRR, \ptwoRB$ and $\pthreeBB, \pthreeBR, \pthreeRR, \pthreeRB$, respectively. Further, let $n(\calRed_\timeValue), n(\calBlue_\timeValue)$ denote the number of nodes at time $\timeValue$ in the red and blue groups respectively and $\groupSize_\timeValue
= n(\calRed_\timeValue) + n(\calBlue_\timeValue)$.

\vspace{0.2cm}
With the above notation, we can express the evolution of the total in-degree of red nodes~$\inDegree(\calRed_{\timeValue})$ as follows:
\begin{align}
\label{eq:proof_degree_i_evolution}
\mathbb{E}\{\inDegree({\calRed_{\timeValue + 1})} - \inDegree({\calRed_{\timeValue})} |
\graph_\timeValue \} &= \probEventOne(\redBirthProb\poneRR + (1-\redBirthProb)\poneBR) + \probEventTwo\redBirthProb + (1-\probEventOne - \probEventTwo)(\pthreeRR + \pthreeBR).
\end{align}
The idea behind (\ref{eq:proof_degree_i_evolution}) is that the total in-degree of red group can increase by an amount of $1$ at time $\timeValue$~(i.e.,~$\inDegree({\calRed_{\timeValue + 1})} - \inDegree({\calRed_{\timeValue})} = 1$) via one of the three ways: event~1 takes place and either a new red node is added and follows an existing red node or a new blue node is added and follows an existing red node, event~2 takes place and a new red node is born or, event~3 takes place and an an existing red node follows an existing blue or a red node. Eq.~(\ref{eq:proof_degree_i_evolution}) expresses the expectation of this event where the three summands corresponds to the three ways in which the in-degree of the red-group increase by one. Since the number of edges in the network $\graph_\timeValue$ at time $\timeValue$ is equal to $\timeValue$~(i.e.,~$|\edgeSet_\timeValue| = \timeValue$), from (\ref{eq:proof_degree_i_evolution}) i) we get,
\begin{align}
&\mathbb{E}\left\{\frac{\inDegree\left(\calRed_{\timeValue + 1}\right)}{\timeValue + 1}|
\graph_\timeValue \right\} =\mathbb{E}\left\{{\thetaIn_{\timeValue + 1}}|
\graph_\timeValue \right\} \nonumber\\ &\hspace{0.5cm}=\frac{\inDegree({\calRed_{\timeValue})}}{\timeValue + 1} +  \frac{1}{\timeValue + 1}\left(\probEventOne(\redBirthProb\poneRR + (1-\redBirthProb)\poneBR) + \probEventTwo\redBirthProb + (1-\probEventOne - \probEventTwo)(\pthreeRR + \pthreeBR) \right) \nonumber\\
&\hspace{0.5cm}= \thetaIn_\timeValue + \frac{1}{\timeValue + 1}\left(\probEventOne(\redBirthProb\poneRR + (1-\redBirthProb)\poneBR) + \probEventTwo\redBirthProb + (1-\probEventOne - \probEventTwo)(\pthreeRR + \pthreeBR) -\thetaIn_\timeValue\right). \label{eq:proof_theta_i_evolution}
\end{align}

Using similar arguments for the total out-degree of red group, we get,
\begin{align}
&\mathbb{E}\left\{\frac{\outDegree\left(\calRed_{\timeValue + 1}\right)}{\timeValue + 1}|
\graph_\timeValue \right\}  = \mathbb{E}\left\{{\thetaOut_{\timeValue + 1}}|
\graph_\timeValue \right\} \nonumber\\ 
&\hspace{0.5cm}= \thetaOut_\timeValue + \frac{1}{\timeValue + 1}\left( \probEventOne\redBirthProb + \probEventTwo(\redBirthProb\ptwoRR + (1-\redBirthProb)\ptwoBR) + (1-\probEventOne - \probEventTwo)(\pthreeRR + \pthreeRB) -\thetaOut_\timeValue\right). \label{eq:proof_theta_o_evolution}
\end{align}

Therefore, by combining (\ref{eq:proof_theta_i_evolution}) and (\ref{eq:proof_theta_o_evolution}), we get:
\begin{align}
&\mathbb{E}\left\{{\thetaVec_{\timeValue + 1}}| \graph_\timeValue\right\}  = \thetaVec_\timeValue + \nonumber\\
&\hspace{0.25cm}\frac{1}{\timeValue + 1}
\begin{bmatrix}
\probEventOne(\redBirthProb\poneRR + (1-\redBirthProb)\poneBR) + \probEventTwo\redBirthProb + (1-\probEventOne - \probEventTwo)(\pthreeRR + \pthreeBR) -\thetaIn_\timeValue  \\
 \probEventOne\redBirthProb + \probEventTwo(\redBirthProb\ptwoRR + (1-\redBirthProb)\ptwoBR) + (1-\probEventOne - \probEventTwo)(\pthreeRR + \pthreeRB) -\thetaOut_\timeValue \label{eq:proof_theta_evolution}
\end{bmatrix}.
\end{align}

Next, let us consider~$\poneRR$. Recall that $\poneRR$ is the probability that, given the event~1 happened and the new node is red, it is followed by an existing red node. This event can happen in three ways: \\
 1~-~a potential red follower is chosen via preferential attachment~(with probability~$\frac{\inDegree(\calRed_{\timeValue}) + \groupSize(\calRed_\timeValue)\deltaboth}{\degree_\timeValue + \groupSize_\timeValue \deltaboth}$) and the new node is followed by it~(probability $\homophilyRedAnyEvent$)
  \\2~-~a potential red follower is chosen via preferential attachment~(with probability~$\frac{\inDegree(\calRed_{\timeValue}) + \groupSize(\calRed_\timeValue)\deltaboth}{\degree_\timeValue + \groupSize_\timeValue \deltaboth}$) and the new node is not followed by it~(probability $1-\homophilyRedAnyEvent$) and then the event takes place after that with probability $\poneRR$\\
 3~-~a potential blue follower is chosen via preferential attachment~(with probability~$\frac{\inDegree(\calBlue_{\timeValue}) + \groupSize(\calBlue_\timeValue)\deltaboth}{\degree_\timeValue + \groupSize_\timeValue \deltaboth}$) and the new node is not followed by it~(probability $\homophilyBlueAnyEvent$) and then the event takes place after that with probability $\poneRR$.

 Hence, $\poneRR$ satisfies,
\begin{align}
\poneRR &= \frac{\inDegree(\calRed_{\timeValue}) + \groupSize(\calRed_\timeValue)}{\degree_\timeValue + \groupSize_\timeValue \deltaboth}\homophilyRedAnyEvent + \left(\frac{\inDegree(\calBlue_{\timeValue}) + \groupSize(\calBlue_\timeValue)\deltaboth}{\degree_\timeValue + \groupSize_\timeValue \deltaboth}\homophilyBlueAnyEvent + \frac{\inDegree(\calRed_{\timeValue}) + \groupSize(\calRed_\timeValue)\deltaboth}{\degree_\timeValue + \groupSize_\timeValue \deltaboth}\left(1-\homophilyRedAnyEvent\right) \right)\poneRR \nonumber\\
 &= \frac{ \left(\inDegree(\calRed_\timeValue) + \groupSize(\calRed_{\timeValue})\deltaboth\right)\homophilyRedAnyEvent}{\degree_\timeValue + \groupSize_\timeValue \deltaboth -\left(		 \left(\inDegree(\calBlue_\timeValue) + \groupSize(\calBlue_{\timeValue})\deltaboth\right)\homophilyBlueAnyEvent	+  \left(\inDegree(\calRed_\timeValue) + \groupSize(\calRed_{\timeValue})\deltaboth\right)\left(1-\homophilyRedAnyEvent\right)\right)} \nonumber\\
 &= \frac{\left(\thetaIn_\timeValue\degree_\timeValue + \groupSize(\calRed_\timeValue)\deltaboth \right)\homophilyRedAnyEvent}{\degree_\timeValue + \groupSize_\timeValue \deltaboth -		 \left( \left(1-\thetaIn_\timeValue\right) \degree_\timeValue + \groupSize(\calBlue_{\timeValue})\deltaboth \right)\homophilyBlueAnyEvent	 -\left( \thetaIn_\timeValue \degree_\timeValue+ \groupSize(\calRed_{\timeValue})\deltaboth\right)\left(1-\homophilyRedAnyEvent\right)} \quad \text{(by definition of $\thetaIn_\timeValue$)} \nonumber\\
 &= \frac{\left(\thetaIn_\timeValue\timeValue + \groupSize(\calRed_\timeValue)\deltaboth \right)\homophilyRedAnyEvent}{\timeValue + \groupSize_\timeValue \deltaboth -		 \left( \left(1-\thetaIn_\timeValue\right) \timeValue + \groupSize(\calBlue_{\timeValue})\deltaboth \right)\homophilyBlueAnyEvent	 -\left( \thetaIn_\timeValue \timeValue+ \groupSize(\calRed_{\timeValue})\deltaboth\right)\left(1-\homophilyRedAnyEvent\right)} \quad \text{(since $\degree_\timeValue = \timeValue$)} \label{eq:poneRR}
\end{align}
Next, by Hoeffding's inequality, we get,
\begin{equation}
\label{eq:hoeffding_poneRR}
\begin{aligned}
\mathbb{P}\left\{	\left|\groupSize\left(\calRed_{\timeValue} \right) -(\probEventOne + \probEventTwo)\redBirthProb\timeValue \right|	\geq \sqrt{\frac{\timeValue \log \timeValue}{2}}\right\} &\leq \frac{1}{\timeValue}\\
\mathbb{P}\left\{	\left|\groupSize\left(\calBlue_{\timeValue} \right) -(\probEventOne + \probEventTwo)\left(1-\redBirthProb\right)\timeValue \right|	\geq \sqrt{\frac{\timeValue \log \timeValue}{2}}\right\} &\leq \frac{1}{\timeValue}\\
\mathbb{P}\left\{	\left|\groupSize_\timeValue -(\probEventOne + \probEventTwo)\timeValue \right|	\geq \sqrt{\frac{\timeValue \log \timeValue}{2}}\right\} &\leq \frac{1}{\timeValue}.
\end{aligned}
\end{equation}
Hence, by (\ref{eq:poneRR}) and (\ref{eq:hoeffding_poneRR}), we observe that, with probability $1-o(\frac{1}{\timeValue})$,
\begin{align}
\poneRR&= \frac{\left(\thetaIn_\timeValue + \left(\probEventOne+\probEventTwo\right)\redBirthProb\deltaboth \right)\homophilyRedAnyEvent}{1 + \left(\probEventOne+\probEventTwo\right)\deltaboth -		 \left( \left(1-\thetaIn_\timeValue\right) + \left(\probEventOne+\probEventTwo\right)\left(1-\redBirthProb\right)\deltaboth \right)\homophilyBlueAnyEvent	 -\left( \thetaIn_\timeValue+ \left(\probEventOne+\probEventTwo\right)\redBirthProb\deltaboth\right)\left(1-\homophilyRedAnyEvent\right)} + O\left(\frac{1}{\timeValue^{\frac{1}{4}}}\right). \nonumber\\
\end{align}
Following similar steps, we also get, with probability $1-o(\frac{1}{\timeValue})$,
\begin{align}
\poneBR&= \frac{\left(\thetaIn_\timeValue + \left(\probEventOne+\probEventTwo\right)\redBirthProb\deltaboth \right)\left(1-\homophilyRedAnyEvent\right)}{1 + \left(\probEventOne+\probEventTwo\right)\deltaboth -		 \left( \left(1-\thetaIn_\timeValue\right) + \left(\probEventOne+\probEventTwo\right)\left(1-\redBirthProb\right)\deltaboth \right)\left(1-\homophilyBlueAnyEvent\right) -\left( \thetaIn_\timeValue+ \left(\probEventOne+\probEventTwo\right)\redBirthProb\deltaboth\right)\homophilyRedAnyEvent} + O\left(\frac{1}{\timeValue^{\frac{1}{4}}}\right). \nonumber\\
\ptwoRR&= \frac{\left(\thetaOut_\timeValue + \left(\probEventOne+\probEventTwo\right)\redBirthProb\deltaboth \right)\homophilyRedAnyEvent}{1 + \left(\probEventOne+\probEventTwo\right)\deltaboth -		 \left( \left(1-\thetaOut_\timeValue\right) + \left(\probEventOne+\probEventTwo\right)\left(1-\redBirthProb\right)\deltaboth \right)\homophilyRedAnyEvent	 -\left( \thetaOut_\timeValue+ \left(\probEventOne+\probEventTwo\right)\redBirthProb\deltaboth\right)\left(1-\homophilyRedAnyEvent\right)} + O\left(\frac{1}{\timeValue^{\frac{1}{4}}}\right). \nonumber\\
\ptwoBR&= \frac{\left(\thetaOut_\timeValue + \left(\probEventOne+\probEventTwo\right)\redBirthProb\deltaboth \right)\left(1-\homophilyBlueAnyEvent\right)}{1 + \left(\probEventOne+\probEventTwo\right)\deltaboth -		 \left( \left(1-\thetaOut_\timeValue\right) + \left(\probEventOne+\probEventTwo\right)\left(1-\redBirthProb\right)\deltaboth \right)\left(1-\homophilyBlueAnyEvent\right)	 -\left( \thetaOut_\timeValue+ \left(\probEventOne+\probEventTwo\right)\redBirthProb\deltaboth\right)\homophilyBlueAnyEvent} + O\left(\frac{1}{\timeValue^{\frac{1}{4}}}\right). \nonumber
\end{align} and,
\begin{align}
\pthreeRR&= \frac{\left(\thetaOut_\timeValue + \left(\probEventOne + \probEventTwo\right)\redBirthProb\deltaboth \right)\left(\thetaIn_\timeValue + \left(\probEventOne + \probEventTwo\right)\redBirthProb\deltaboth \right)\homophilyRedAnyEvent}{
	\splitfrac{\left(1+\left(\probEventOne+\probEventTwo\right)\deltaboth\right)^2 }{
	\splitfrac{
	 - \left( 1	-\thetaOut_\timeValue+ \left(\probEventOne+\probEventTwo\right)\left(1-\redBirthProb\right)\deltaboth\right)\left( 1	-\thetaIn_\timeValue+ \left(\probEventOne+\probEventTwo\right)\left(1-\redBirthProb\right)\deltaboth\right)\left(1-\homophilyBlueAnyEvent\right)}{\splitfrac{- \left( 1	-\thetaOut_\timeValue+ \left(\probEventOne+\probEventTwo\right)\left(1-\redBirthProb\right)\deltaboth\right)\left(\thetaIn_\timeValue+ \left(\probEventOne+\probEventTwo\right)\redBirthProb\deltaboth\right)\homophilyRedAnyEvent }{\splitfrac{- \left( \thetaOut_\timeValue+ \left(\probEventOne+\probEventTwo\right)\redBirthProb\deltaboth\right)\left( 1	-\thetaIn_\timeValue+ \left(\probEventOne+\probEventTwo\right)\left(1-\redBirthProb\right)\deltaboth\right)\homophilyBlueAnyEvent}{
	 	- \left(\thetaOut_\timeValue+ \left(\probEventOne+\probEventTwo\right)\redBirthProb\deltaboth\right)\left(	\thetaIn_\timeValue+ \left(\probEventOne+\probEventTwo\right)\redBirthProb\deltaboth\right)\left(1-\homophilyRedAnyEvent\right)
 	}	 
 } 
}
}
}	+ O\left(\frac{1}{\timeValue^{\frac{1}{4}}}\right)	\nonumber\\
\pthreeBR&= \frac{\left(1-\thetaOut_\timeValue + \left(\probEventOne + \probEventTwo\right)\left(1-\redBirthProb\right)\deltaboth \right)\left(\thetaIn_\timeValue + \left(\probEventOne + \probEventTwo\right)\redBirthProb\deltaboth \right)\left(1-\homophilyRedAnyEvent\right)}{
	\splitfrac{\left(1+\left(\probEventOne+\probEventTwo\right)\deltaboth\right)^2 }{
		\splitfrac{
			- \left( 1	-\thetaOut_\timeValue+ \left(\probEventOne+\probEventTwo\right)\left(1-\redBirthProb\right)\deltaboth\right)\left( 1	-\thetaIn_\timeValue+ \left(\probEventOne+\probEventTwo\right)\left(1-\redBirthProb\right)\deltaboth\right)\left(1-\homophilyBlueAnyEvent\right)}{\splitfrac{- \left( 1	-\thetaOut_\timeValue+ \left(\probEventOne+\probEventTwo\right)\left(1-\redBirthProb\right)\deltaboth\right)\left(\thetaIn_\timeValue+ \left(\probEventOne+\probEventTwo\right)\redBirthProb\deltaboth\right)\homophilyRedAnyEvent }{\splitfrac{- \left( \thetaOut_\timeValue+ \left(\probEventOne+\probEventTwo\right)\redBirthProb\deltaboth\right)\left( 1	-\thetaIn_\timeValue+ \left(\probEventOne+\probEventTwo\right)\left(1-\redBirthProb\right)\deltaboth\right)\homophilyBlueAnyEvent}{
					- \left(\thetaOut_\timeValue+ \left(\probEventOne+\probEventTwo\right)\redBirthProb\deltaboth\right)\left(	\thetaIn_\timeValue+ \left(\probEventOne+\probEventTwo\right)\redBirthProb\deltaboth\right)\left(1-\homophilyRedAnyEvent\right)
				}	 
			} 
		}
	}
}	+ O\left(\frac{1}{\timeValue^{\frac{1}{4}}}\right)	\nonumber\\
\pthreeRB&= \frac{\left(\thetaOut_\timeValue + \left(\probEventOne + \probEventTwo\right)\redBirthProb\deltaboth \right)\left(1-\thetaIn_\timeValue + \left(\probEventOne + \probEventTwo\right)\left(1-\redBirthProb\right)\deltaboth \right)\left(1-\homophilyBlueAnyEvent\right)}{
	\splitfrac{\left(1+\left(\probEventOne+\probEventTwo\right)\deltaboth\right)^2 }{
		\splitfrac{
			- \left( 1	-\thetaOut_\timeValue+ \left(\probEventOne+\probEventTwo\right)\left(1-\redBirthProb\right)\deltaboth\right)\left( 1	-\thetaIn_\timeValue+ \left(\probEventOne+\probEventTwo\right)\left(1-\redBirthProb\right)\deltaboth\right)\left(1-\homophilyBlueAnyEvent\right)}{\splitfrac{- \left( 1	-\thetaOut_\timeValue+ \left(\probEventOne+\probEventTwo\right)\left(1-\redBirthProb\right)\deltaboth\right)\left(\thetaIn_\timeValue+ \left(\probEventOne+\probEventTwo\right)\redBirthProb\deltaboth\right)\homophilyRedAnyEvent }{\splitfrac{- \left( \thetaOut_\timeValue+ \left(\probEventOne+\probEventTwo\right)\redBirthProb\deltaboth\right)\left( 1	-\thetaIn_\timeValue+ \left(\probEventOne+\probEventTwo\right)\left(1-\redBirthProb\right)\deltaboth\right)\homophilyBlueAnyEvent}{
					- \left(\thetaOut_\timeValue+ \left(\probEventOne+\probEventTwo\right)\redBirthProb\deltaboth\right)\left(	\thetaIn_\timeValue+ \left(\probEventOne+\probEventTwo\right)\redBirthProb\deltaboth\right)\left(1-\homophilyRedAnyEvent\right)
				}	 
			} 
		}
	}
}	+ O\left(\frac{1}{\timeValue^{\frac{1}{4}}}\right)	\nonumber
\end{align}

Now, we have all the ingredients we need to show that the assumptions C.1-C.6 are satisfied Theorem~\ref{th:kushner_theorem} for the sequence $\{\thetaVec_\timeValue\}$. 

First, note that $\{\thetaVec_\timeValue\}$ is Markovian implying that its evolution can be expressed in the form (\ref{eq:algorithm_kushner_theorem}) with $z_\timeValue = 0$ for all $\timeValue = 1, 2, \dots$ since $\thetaVec_\timeValue$ is naturally in the space $[0,1]^2$~i.e.,~$\thetaVec_{\timeValue + 1} = \thetaVec_\timeValue + \epsilon_{\timeValue}Y_\timeValue$. This further implies that C.1 is satisfied. 

Next, note that some of the terms on the right side of (\ref{eq:proof_theta_evolution}) are constant model parameters~($\redBirthProb$, $\probEventOne$, $\probEventTwo$, $\homophilyBlueAnyEvent$, $\homophilyRedAnyEvent$, $\deltaboth$) and each of the remaining terms~($\poneRR$, $\poneBR$, $\ptwoRR$, $\ptwoBR$, $\pthreeRR$, $\pthreeBR$, $\pthreeRB$) is equal to a function of $\thetaIn_\timeValue, \thetaOut_\timeValue$ plus a noise term that is $O\left(\frac{1}{\timeValue^{{1}/{4}}}\right)$ with probability $1-o\left(\frac{1}{\timeValue}\right)$. Further, we can see from (\ref{eq:proof_theta_evolution}) that $\epsilon_{\timeValue} = \frac{1}{\timeValue + 1}$ and since the noise term is $O\left(\frac{1}{\timeValue^{{1}/{4}}}\right)$, C.6 is satisfied. Further, we get,
\begin{equation}
\label{eq:proof_F_expression}
\begin{aligned}
g(\thetaVec_{\timeValue}) &= \begin{bmatrix}
\probEventOne(\redBirthProb\barponeRR + (1-\redBirthProb)\barponeBR) + \probEventTwo\redBirthProb + (1-\probEventOne - \probEventTwo)(\barpthreeRR + \barpthreeBR) \\
\probEventOne\redBirthProb + \probEventTwo(\redBirthProb\barptwoRR + (1-\redBirthProb)\barptwoBR) + (1-\probEventOne - \probEventTwo)(\barpthreeRR + \barpthreeRB) 
\end{bmatrix} - \begin{bmatrix}
\thetaIn_\timeValue \\
\thetaOut_\timeValue 
\end{bmatrix}\\
&=\nonLinFunction(\thetaVec_\timeValue) - \thetaVec_{\timeValue}
\end{aligned}
\end{equation}
where, $\barponeRR$, $\barponeBR$, $\barptwoRR$, $\barptwoBR$, $\barpthreeRR$, $\barpthreeBR$, $\barpthreeRB$ correspond to $\poneRR$, $\poneBR$, $\ptwoRR$, $\ptwoBR$, $\pthreeRR$, $\pthreeBR$, $\pthreeRB$, respectively, with the $O\left(\frac{1}{\timeValue^{{1}/{4}}}\right)$ noise term neglected. 

Hence, the only remaining task left is to show that $\nonLinFunction(\cdot)$ (shown in (\ref{eq:proof_F_expression})) is a contraction map. For this, after some tedious algebra, we get the Jacobian $J$ of $\nonLinFunction(\cdot)$ to be of a special form with respect to $\deltaboth$: each element (denoted by~$J_{ij}\, i,j = 1,2$) of $J$ is a ratio of two polynomials of $\deltaboth$ where the denominator polynomial is the higher order one. This implies that there exists $\deltaboth^{*} >0$ such that, for all $\deltaboth > \deltaboth^{*}$,
\begin{align}
\left|\left|J\right|\right|_2 \leq \left|\left|J\right|\right|_F = \left(\sum_{i}\sum_{j}\left|J_{ij}\right|^2\right)^\frac{1}{2} < 1
\end{align}
where $\left|\left|\cdot\right|\right|_F$ denotes the Frobenius norm and $\left|\left|\cdot\right|\right|_2$ denotes the  largest singular value~(spectral norm). Hence, there exists $\deltaboth^{*} >0$ such that, for all $\deltaboth > \deltaboth^{*}$, $\nonLinFunction(\cdot)$ is a contraction map.

\subsection{Proof of part~ii of Theorem~\ref{th:convergence_DMPA}}
We will provide the proof for the in-degree distribution. The proof for the out-degree distribution involves similar arguments. 

\vspace{0.1cm}
\noindent
{\bf Notation:} We first need to establish some additional notation that we need for the proof of second part. Let,
\begin{compactitem}
		\item[] $\pOneBRk : $ given the event~$1$ and the new node is blue, probability that it is followed by an existing red node with in-degree $k$
		\item[] $\pOneRRk: $ given the event~$1$ and the new node is red, probability that it is followed by an existing red node with in-degree $k$
		\item[] $\pThreeBRk : $ given the event~$3$, probability that  an existing blue node is followed by an existing red node with in-degree $k$.
		\item[] $\pThreeRRk : $ given the event~$3$, probability that  an existing red node is followed by an existing red node with in-degree $k$.		
\end{compactitem}	
Now, let us consider the evolution of the number of $\numNodesIndegreeTime(\calRed_\timeValue)$ which is the number of red nodes with in-degree $k$ at time $\timeValue$ as defined in (\ref{eq:numNodesDegreeTime}). Note that, for any $k = 1,2,\dots$, we can write the evolution of $\numNodesIndegreeTime(\calRed_\timeValue)$  as,
\begin{equation}
\label{eq:proof_evolution_in_degree}
\begin{aligned}
\mathbb{E}\left\{	\numNodesIndegreeTime\left(\calRed_{\timeValue+1}\right) \big| \graph_\timeValue		\right\} &= 	\numNodesIndegreeTime\left(\calRed_{\timeValue}\right) +\mathbb{P}\left\{	\text{a red node with in-degree $k-1$ gets a friend}	\right\} \\
&\hspace{0.2cm}- \mathbb{P}\left\{	\text{a red node with in-degree $k$ gets a friend}	\right\} + \probEventOne\redBirthProb\mathds{1}_{\left\{k=0\right\}} +  \probEventTwo\redBirthProb\mathds{1}_{\left\{k=1\right\}}.\\
\end{aligned}
\end{equation}

The idea behind (\ref{eq:proof_evolution_in_degree}) is as follows: the number of red nodes with in-degree $k$ at time $\timeValue$ can change in two ways for $k>1$: a red node with in-degree $k-1$ follows another node and thus $\numNodesIndegreeTime(\calRed_\timeValue)$ increase by $1$ or, a red node with in-degree $k$ follows another node and thus $\numNodesIndegreeTime(\calRed_\timeValue)$ decrease by $1$. In addition, the number of red nodes with in-degree $k=0$ can increase when the event~1 happens and the number of red nodes with in-degree $k=1$ can increase by $1$ when the event~2 happens - these situations are captured by the last two summands in (\ref{eq:proof_evolution_in_degree}). Next, note that,
\begin{equation}
\label{eq:proof_red_k_in_degree_gets_a_friend}
\begin{aligned}
\mathbb{P}\left\{	\text{a red node with in-degree $k$ gets a friend}	\right\} &= \probEventOne\left(	\left(	1-\redBirthProb\right)\pOneBRk + \redBirthProb\pOneRRk	\right) \\ &\hspace{0.23cm}+ \left(1-\probEventOne-\probEventTwo\right)\left(\pThreeBRk + \pThreeRRk\right)
\end{aligned}
\end{equation}
where, the idea is that a red node with in-degree $k$ can follow a blue or a red new node in the event~1 or, an existing red node with in-degree $k$ can follow an existing blue or a red node in the event~3. 

Next, let us consider, $\pOneBRk$ which is the probability that, given the event~1 happened and the new node is blue, a red node with in-degree $k$ follows it. This can happen in three ways:\\
1~-~a potential red follower with in-degree $k$ chosen via preferential attachment~(with probability~$\numNodesIndegreeTime\left(\calRed_{\timeValue}\right)\left(\frac{k + \deltaboth}{\degree_\timeValue + \groupSize_\timeValue \deltaboth}\right)$) and the new blue node is followed by it~(probability $1-\homophilyRedAnyEvent$)
\\2~-~a potential red follower is chosen via preferential attachment~(with probability~$\frac{\inDegree\left(\calRed_{\timeValue}\right) + \groupSize\left(\calRed_{\timeValue}\right)\deltaboth}{\degree_\timeValue + \groupSize_\timeValue \deltaboth}$) and the new blue node is not followed by it~(probability $\homophilyRedAnyEvent$) and then the event takes place after that\\
3~-~a potential blue follower is chosen via preferential attachment~(with probability~$\frac{\inDegree\left(\calBlue_{\timeValue}\right) + \groupSize\left(\calBlue_{\timeValue}\right)\deltaboth}{\degree_\timeValue + \groupSize_\timeValue \deltaboth}$) and the new blue node is not followed by it~(probability $1-\homophilyBlueAnyEvent$) and then the event takes place after that.

Hence, $\pOneBRk$ satisfies,
\begin{align}
\pOneBRk &= \numNodesIndegreeTime\left(\calRed_{\timeValue}\right)\left(\frac{k + \deltaboth}{\degree_\timeValue + \groupSize_\timeValue \deltaboth}\right)\left(1-\homophilyRedAnyEvent\right) \nonumber\\ &\hspace{1cm}+  \left( \frac{\inDegree\left(\calRed_{\timeValue}\right) + \groupSize\left(\calRed_{\timeValue}\right)\deltaboth}{\degree_\timeValue + \groupSize_\timeValue \deltaboth}\homophilyRedAnyEvent +  \frac{\inDegree\left(\calBlue_{\timeValue}\right) + \groupSize\left(\calBlue_{\timeValue}\right)\deltaboth}{\degree_\timeValue + \groupSize_\timeValue \deltaboth}\left(1-\homophilyBlueAnyEvent\right) \right)\pOneBRk\nonumber\\
&= \frac{\numNodesIndegreeTime\left(\calRed_{\timeValue}\right)}{1-\left( \frac{\inDegree\left(\calRed_{\timeValue}\right) + \groupSize\left(\calRed_{\timeValue}\right)\deltaboth}{\degree_\timeValue + \groupSize_\timeValue \deltaboth}\homophilyRedAnyEvent +  \frac{\inDegree\left(\calBlue_{\timeValue}\right) + \groupSize\left(\calBlue_{\timeValue}\right)\deltaboth}{\degree_\timeValue + \groupSize_\timeValue \deltaboth}\left(1-\homophilyBlueAnyEvent\right) \right)}\left(\frac{k + \deltaboth}{\degree_\timeValue + \groupSize_\timeValue \deltaboth}\right)\left(1-\homophilyRedAnyEvent\right)
\end{align}
Following similar arguments, we also get,
\begin{align}
\pOneRRk &= \frac{\numNodesIndegreeTime\left(\calRed_{\timeValue}\right)}{1-\left( \frac{\inDegree\left(\calRed_{\timeValue}\right) + \groupSize\left(\calRed_{\timeValue}\right)\deltaboth}{\degree_\timeValue + \groupSize_\timeValue \deltaboth}\left(1-\homophilyRedAnyEvent\right) +  \frac{\inDegree\left(\calBlue_{\timeValue}\right) + \groupSize\left(\calBlue_{\timeValue}\right)\deltaboth}{\degree_\timeValue + \groupSize_\timeValue \deltaboth}\homophilyBlueAnyEvent\right)}\left(\frac{k + \deltaboth}{\degree_\timeValue + \groupSize_\timeValue \deltaboth}\right)\homophilyRedAnyEvent 
\end{align}
\begin{align}
\pThreeBRk = \frac{\numNodesIndegreeTime\left(\calRed_{\timeValue}\right) \left(\outDegree\left(\calBlue_{\timeValue}\right)+\groupSize\left(\calBlue_\timeValue\right)\deltaboth\right) \left(k + \deltaboth\right)\left(1-\homophilyRedAnyEvent\right)}{\splitfrac{\left(\degree_\timeValue +\groupSize_\timeValue\deltaboth\right)^2 - \left(\outDegree\left(\calBlue_{\timeValue}\right)+\groupSize\left(\calBlue_\timeValue\right)\deltaboth\right)\left(\inDegree\left(\calBlue_{\timeValue}\right)+\groupSize\left(\calBlue_\timeValue\right)\deltaboth\right)\left(1-\homophilyBlueAnyEvent\right)}{\splitdfrac{-\left(\outDegree\left(\calBlue_{\timeValue}\right)+\groupSize\left(\calBlue_\timeValue\right)\deltaboth\right)\left(\inDegree\left(\calRed_{\timeValue}\right)+\groupSize\left(\calRed_\timeValue\right)\deltaboth\right)\homophilyRedAnyEvent}{\splitfrac{-\left(\outDegree\left(\calRed_{\timeValue}\right)+\groupSize\left(\calRed_\timeValue\right)\deltaboth\right)\left(\inDegree\left(\calRed_{\timeValue}\right)+\groupSize\left(\calRed_\timeValue\right)\deltaboth\right)\left(1-\homophilyRedAnyEvent\right)}
			{-\left(\outDegree\left(\calRed_{\timeValue}\right)+\groupSize\left(\calRed_\timeValue\right)\deltaboth\right)\left(\inDegree\left(\calBlue_{\timeValue}\right)+\groupSize\left(\calBlue_\timeValue\right)\deltaboth\right)\homophilyRedAnyEvent}
		}
	}
}\\
\pThreeRRk = \frac{\numNodesIndegreeTime\left(\calRed_{\timeValue}\right) \left(\outDegree\left(\calRed_{\timeValue}\right)+\groupSize\left(\calRed_\timeValue\right)\deltaboth\right) \left(k + \deltaboth\right)\homophilyRedAnyEvent}{\splitfrac{\left(\degree_\timeValue +\groupSize_\timeValue\deltaboth\right)^2 - \left(\outDegree\left(\calBlue_{\timeValue}\right)+\groupSize\left(\calBlue_\timeValue\right)\deltaboth\right)\left(\inDegree\left(\calBlue_{\timeValue}\right)+\groupSize\left(\calBlue_\timeValue\right)\deltaboth\right)\left(1-\homophilyBlueAnyEvent\right)}{\splitdfrac{-\left(\outDegree\left(\calBlue_{\timeValue}\right)+\groupSize\left(\calBlue_\timeValue\right)\deltaboth\right)\left(\inDegree\left(\calRed_{\timeValue}\right)+\groupSize\left(\calRed_\timeValue\right)\deltaboth\right)\homophilyRedAnyEvent}{\splitfrac{-\left(\outDegree\left(\calRed_{\timeValue}\right)+\groupSize\left(\calRed_\timeValue\right)\deltaboth\right)\left(\inDegree\left(\calRed_{\timeValue}\right)+\groupSize\left(\calRed_\timeValue\right)\deltaboth\right)\left(1-\homophilyRedAnyEvent\right)}
			{-\left(\outDegree\left(\calRed_{\timeValue}\right)+\groupSize\left(\calRed_\timeValue\right)\deltaboth\right)\left(\inDegree\left(\calBlue_{\timeValue}\right)+\groupSize\left(\calBlue_\timeValue\right)\deltaboth\right)\homophilyRedAnyEvent}
		}
	}
}.
\end{align}
Then, by substituting using the expressions for $\pOneBRk$, $\pOneRRk$, $\pThreeBRk$, $\pThreeRRk$ in (\ref{eq:proof_evolution_in_degree}) and using Hoeffding's inequality to replace $\groupSize_\timeValue, \groupSize(\calBlue_\timeValue), \groupSize(\calRed_\timeValue)$ with their expectations~(similar to the proof of part~i), we get, with probability $1-o\left(\frac{1}{\timeValue}\right)$,
\begin{equation}
\label{eq:proof_evolution_in_degree_final}
\begin{aligned}
\mathbb{E}\left\{\frac{\numNodesIndegreeTime\left(\calRed_{\timeValue+1}\right)}{\timeValue+1} \Bigg| \graph_\timeValue		\right\} &=	\frac{\numNodesIndegreeTime\left(\calRed_\timeValue\right)}{\timeValue} \\
&\hspace{-1.5cm}-  \frac{1}{\timeValue+1}\left(	\left(k+\deltaboth\right)\A\left(\thetaVec_{\timeValue}\right)\frac{\numNodesIndegreeTime\left(\calRed_\timeValue\right)}{\timeValue}	- \left(k-1+\deltaboth\right)\A\left(\thetaVec_{\timeValue}\right)\frac{\numNodesIndegreeminusoneTime\left(\calRed_\timeValue\right)}{\timeValue} + \frac{\numNodesIndegreeTime\left(\calRed_\timeValue\right)}{\timeValue}	 +O\left(\frac{1}{\timeValue^{1/4}}\right)\right)
\end{aligned}
\end{equation}
where,
\begin{equation}
\label{eq:proof_constant_A_in}
\begin{aligned}
\A(\thetaVec_{\timeValue}) &= \frac{\probEventOne\left(1-\redBirthProb\right)\left(1-\homophilyRedAnyEvent\right)}{1 + \deltaboth\left(\probEventOne+ \probEventTwo\right)  -\left(\thetaIn_\timeValue + \deltaboth\left(\probEventOne+\probEventTwo\right)\redBirthProb \right)\homophilyRedAnyEvent - \left(1-\thetaIn_\timeValue + \deltaboth\left(\probEventOne+\probEventTwo\right)\left(1-\redBirthProb\right) \right)\left(1-\homophilyBlueAnyEvent\right)}\\
&+ \frac{\probEventOne\redBirthProb\homophilyRedAnyEvent}{1 + \deltaboth\left(\probEventOne+ \probEventTwo\right)  -\left(\thetaIn_\timeValue + \deltaboth\left(\probEventOne+\probEventTwo\right)\redBirthProb \right)\left(1-\homophilyRedAnyEvent\right) - \left(1-\thetaIn_\timeValue + \deltaboth\left(\probEventOne+\probEventTwo\right)\left(1-\redBirthProb\right) \right)\homophilyBlueAnyEvent}\\
&+\frac{\left(1-\probEventOne-\probEventTwo\right)\left( \left( 1-\thetaOut_\timeValue+\left(\probEventOne+\probEventTwo\right)\left(1-\redBirthProb\right)\deltaboth \right) \left(1-\homophilyRedAnyEvent\right)  + \left(\thetaOut_\timeValue + \left(\probEventOne+\probEventTwo\right)\redBirthProb\deltaboth\right)\homophilyRedAnyEvent  			\right)}{
	\splitfrac{\left(1+\left(\probEventOne+\probEventTwo\right)\deltaboth\right)^2 }{
		\splitfrac{
			- \left( 1	-\thetaOut_\timeValue+ \left(\probEventOne+\probEventTwo\right)\left(1-\redBirthProb\right)\deltaboth\right)\left( 1	-\thetaIn_\timeValue+ \left(\probEventOne+\probEventTwo\right)\left(1-\redBirthProb\right)\deltaboth\right)\left(1-\homophilyBlueAnyEvent\right)}{\splitfrac{- \left( 1	-\thetaOut_\timeValue+ \left(\probEventOne+\probEventTwo\right)\left(1-\redBirthProb\right)\deltaboth\right)\left(\thetaIn_\timeValue+ \left(\probEventOne+\probEventTwo\right)\redBirthProb\deltaboth\right)\homophilyRedAnyEvent }{\splitfrac{- \left( \thetaOut_\timeValue+ \left(\probEventOne+\probEventTwo\right)\redBirthProb\deltaboth\right)\left( 1	-\thetaIn_\timeValue+ \left(\probEventOne+\probEventTwo\right)\left(1-\redBirthProb\right)\deltaboth\right)\homophilyBlueAnyEvent}{
					- \left(\thetaOut_\timeValue+ \left(\probEventOne+\probEventTwo\right)\redBirthProb\deltaboth\right)\left(	\thetaIn_\timeValue+ \left(\probEventOne+\probEventTwo\right)\redBirthProb\deltaboth\right)\left(1-\homophilyRedAnyEvent\right)
				}	 
			} 
		}
	}
}
\end{aligned}.
\end{equation}
Note that (\ref{eq:proof_evolution_in_degree_final}) indicates that the evolution of $\frac{\numNodesIndegreeTime\left(\calRed_\timeValue\right)}{\timeValue}$ satisfies the assumptions in Theorem~\ref{th:kushner_theorem}. More specifically, we can see that the evolution of $\frac{\numNodesIndegreeTime\left(\calRed_\timeValue\right)}{\timeValue}$ is of the form,
\begin{equation*}
\begin{aligned}
\mathbb{E}\left\{	\frac{\numNodesIndegreeTime\left(\calRed_{\timeValue+1}\right) }{\timeValue+1}	\Bigg| \graph_\timeValue	\right\} =	\frac{\numNodesIndegreeTime\left(\calRed_\timeValue\right)}{\timeValue} -  \frac{1}{\timeValue+1}\left(	H\left(\frac{\numNodesIndegreeTime\left(\calRed_\timeValue\right)}{\timeValue}, \frac{\numNodesIndegreeminusoneTime\left(\calRed_\timeValue\right)}{\timeValue}, \A\left(\thetaVec_{\timeValue}\right)\right) + O\left(\frac{1}{\timeValue^{1/4}}\right) \right)
\end{aligned}
\end{equation*}
where the function $H(\cdot)$ is,
\begin{equation*}
H\left(\frac{\numNodesIndegreeTime\left(\calRed_\timeValue\right)}{\timeValue}, \frac{\numNodesIndegreeminusoneTime\left(\calRed_\timeValue\right)}{\timeValue}, \A\left(\thetaVec_{\timeValue}\right)\right) = \left(k+\deltaboth\right)\A\left(\thetaVec_{\timeValue}\right)\frac{\numNodesIndegreeTime\left(\calRed_\timeValue\right)}{\timeValue}	- \left(k-1+\deltaboth\right)\A\left(\thetaVec_{\timeValue}\right)\frac{\numNodesIndegreeminusoneTime\left(\calRed_\timeValue\right)}{\timeValue} + \frac{\numNodesIndegreeTime\left(\calRed_\timeValue\right)}{\timeValue}.	
\end{equation*}
Next, let us consider the autonomous system,
\begin{equation}
\label{eq:proof_ODE_in_degree}
\frac{d}{d\timeValue}\frac{\numNodesIndegreeTime\left(\calRed_\timeValue\right) }{\timeValue}= H\left(\frac{\numNodesIndegreeTime\left(\calRed_\timeValue\right)}{\timeValue}, \frac{\numNodesIndegreeminusoneTime\left(\calRed_\timeValue\right)}{\timeValue}, \A\left(\thetaVec_{\timeValue}\right)\right)
\end{equation}
which can be thought of as the deterministic ODE that approximates the stochastic dynamics of~$\frac{\numNodesIndegreeTime\left(\calRed_{\timeValue+1}\right)}{\timeValue}$. Now recall that $\thetaVec_{\timeValue} = \left[\thetaIn_\timeValue, \thetaOut_\timeValue\right]$ converges almost surely to $\thetaVec_{*} = \left[\thetaIn_{*}, \thetaOut_{*}\right]^T$~(from part~i of Theorem~\ref{th:convergence_DMPA}) and therefore, $\A\left(\thetaVec_{\timeValue}\right)$ converges almost surely to $\A\left(\thetaVec_{*}\right)$. Therefore, setting $H\left(\numNodesIndegreeLimit(\calRed), \numNodesIndegreeminusoneLimit(\calRed),  \A\left(\thetaVec_{*}\right)\right) = 0$~(where, $\numNodesIndegreeLimit(\calRed)$ denotes the equilibrium state of~$\frac{\numNodesIndegreeTime\left(\calRed_{\timeValue+1}\right)}{\timeValue}$), we have,
\begin{align}
\label{eq:proof_limit_ratio_k_kminus1}
\frac{\numNodesIndegreeLimit}{\numNodesIndegreeminusoneLimit} = \frac{\left(k-1+\deltaboth\right)\A\left(\thetaVec_{*}\right) }{\left(k+\deltaboth\right)\A\left(\thetaVec_{*}\right) + 1} =  1 - \left(\frac{1+\frac{1}{\A\left(\thetaVec_{*}\right)}}{k} \right) + O\left(\frac{1}{k^2}\right).
\end{align}
Then, the Lemma~\ref{lemma:power_law_form} and (\ref{eq:proof_limit_ratio_k_kminus1}) imply that the in-degree distribution distribution is a power-law with exponent~i.e.,~${1+\frac{1}{\A\left(\thetaVec_{*}\right)}}$ for large enough $k$~(i.e.,~the tail of the distribution takes power-law form). 

\section{Proof of Theorem~\ref{th:exponents_for_simplified_cases}}
\label{appendix:proof_exponents_for_simplified_cases}

The $\thetaVec_{*} = \left[\thetaIn_{*}, \thetaOut_{*}\right]^T$ for each case can be obtained by substituting the parameter values in $\nonLinFunction(\thetaVec_{*}) - \thetaVec_{*} = 0 $ where $\nonLinFunction(\cdot)$ is given in (\ref{eq:proof_F_expression}). Then, substituting $\thetaIn_{*}, \thetaOut_{*}$ in (\ref{eq:power_law_form}) yields the power-law exponent values. 

\end{document}